\documentclass[12pt]{iopart}

\usepackage{amssymb} %
\usepackage{amsthm} %
\expandafter\let\csname equation*\endcsname\relax %
\expandafter\let\csname endequation*\endcsname\relax %
\usepackage{amsmath} %
\usepackage{color} %
\usepackage{graphicx} %
\usepackage{subfig}

\newcommand\ket[1]{\ensuremath{|#1\rangle}} %
\newcommand\bra[1]{\ensuremath{\langle#1|}} %
\newcommand\argmax{\mathop{\rm argmax}} %
\newcommand\balpha{\boldsymbol{\alpha}} %
\newcommand\brho{\boldsymbol{\rho}} %
\def\C{\mathbb{C}} %
\def\D{\mathcal{D}} %
\def\F{\mathcal{F}} %
\def\H{\mathcal{H}} %
\def\L{\mathcal{L}} %
\def\P{\mathcal{P}} %
\def\R{\mathbb{R}} %

\newtheorem{observation}{Observation} %
\newtheorem{example}{Example} %

\newenvironment{obsprimed}[1]{
  \renewcommand{\theobservation}{\ref{#1}$'$} %
  \addtocounter{observation}{-1} %
  \begin{observation}}{\end{observation}}

\begin{document}


\title{Discontinuity of Maximum Entropy Inference and Quantum Phase
  Transitions}

\author{Jianxin Chen} %
\address{Joint Center for Quantum Information and Computer Science,
  University of Maryland, College Park, Maryland, USA}

\author{Zhengfeng Ji} %
\address{Institute for Quantum Computing, University of Waterloo,
  Waterloo, Ontario, Canada} %
\address{State Key Laboratory of Computer Science, Institute of
  Software, Chinese Academy of Sciences, Beijing, China}

\author{Chi-Kwong Li} %
\address{Department of Mathematics, College of William and Mary,
  Williamsburg, Virginia, USA}

\author{Yiu-Tung Poon} %
\address{Department of Mathematics, Iowa State University, Ames, Iowa,
  USA}

\author{Yi Shen} %
\address{Department of Statistics and Actuarial Science, University of
  Waterloo, Waterloo, Ontario, Canada}

\author{Nengkun Yu} %
\address{Institute for Quantum Computing, University of Waterloo,
  Waterloo, Ontario, Canada} %
\address{Department of Mathematics \& Statistics, University of
  Guelph, Guelph, Ontario, Canada}

\author{Bei Zeng} %
\address{Institute for Quantum Computing, University of Waterloo,
  Waterloo, Ontario, Canada} %
\address{Department of Mathematics \& Statistics, University of
  Guelph, Guelph, Ontario, Canada} %
\address{Canadian Institute for Advanced Research, Toronto, Ontario,
  Canada}

\author{Duanlu Zhou} %
\address{Beijing National Laboratory for Condensed Matter Physics, and
  Institute of Physics, Chinese Academy of Sciences, Beijing 100190,
  China}

\begin{abstract}
  In this paper, we discuss the connection between two genuinely
  quantum phenomena---the discontinuity of quantum maximum entropy
  inference and quantum phase transitions at zero temperature. It is
  shown that the discontinuity of the maximum entropy inference of
  local observable measurements signals the non-local type of
  transitions, where local density matrices of the ground state change
  smoothly at the transition point. We then propose to use the quantum
  conditional mutual information of the ground state as an indicator
  to detect the discontinuity and the non-local type of quantum phase
  transitions in the thermodynamic limit.
\end{abstract}

\section{Introduction}
\label{sec:intro}

Quantum phase transitions happen at zero temperature with no classical
counterparts and are believed to be driven by quantum
fluctuations~\cite{sachdev2007quantum}. The study of quantum phase
transitions has been a central topic in the condensed matter physics
community during the past several decades involving the study of
exotic phases of matter such as
superconductivity~\cite{bardeen1957microscopic}, fractional quantum
Hall systems~\cite{laughlin1983anomalous}, and recently the
topological
insulators~\cite{pankratov1987supersymmetry,bernevig2006quantum,konig2007quantum}.
In recent years, it also becomes an intensively studied topic in
quantum information science community, mainly because of its intimate
connection to the study of local
Hamiltonians~\cite{hastings2010locality}.

In a usual model for quantum phase transitions, one considers a local
Hamiltonian $H(\boldsymbol{\lambda})$ which depends on some parameter
vector $\boldsymbol{\lambda}$. While $H(\boldsymbol{\lambda})$
smoothly changes with $\boldsymbol{\lambda}$, the change of the ground
state $\ket{\psi_0(\boldsymbol{\lambda})}$ may not be smooth when the
system is undergoing a phase transition. Such kind of phenomena is
naturally expected to happen at a level-crossing, or at an avoided
(but near) level-crossing~\cite{sachdev2007quantum}.

Intuitively, the change of ground states can then be measured by some
distance between $\ket{\psi_0(\boldsymbol{\lambda})}$ and
$\ket{\psi_0(\boldsymbol{\lambda}+\delta\boldsymbol{\lambda})}$. For a
small change of the parameters $\boldsymbol{\lambda}$, such a distance
is relatively large near a transition point, while the Hamiltonian
changes smoothly from $H(\boldsymbol{\lambda})$ to
$H(\boldsymbol{\lambda}+\delta\boldsymbol{\lambda})$. The fidelity
approach, using the fidelity of quantum states to measure the change
of the global ground states, has demonstrated the idea successfully in
many physical models for signaling quantum phase
transitions~\cite{zanardi2007information,gu2010fidelity}. While the
fidelity approach is believed to provide a signal for many kinds of
quantum phase transitions, it does not distinguish between different
types of the transition, for instance local or non-local (in a sense
that the reduced fidelity of local density matrices may also signal
the phase transition, as discussed in~\cite{gu2010fidelity}).
Moreover, one usually needs to compute the fidelity change of a
relatively large system in order to clearly signal the transition
point.

In this work, we explore an information-theoretic viewpoint to quantum
phase transitions. Our approach is based on the structure of the
convex set given by all the possible local measurement results, and
the corresponding inference of the global quantum states based on
these local measurement results. By the principle of maximum entropy,
the best such inference compatible with the given local measurement
results is the unique quantum state $\rho^*$ with the maximum von
Neumann entropy~\cite{Jaynes:1957fy}.

It is known that in the classical case, the maximum entropy inference
is continuous~\cite{Jaynes:1957fy,Wichmann:2004gd,knauf2010entropy}.
This means that, for any two sets of local measurement results
${\balpha}$ and ${\balpha}'$ close to each other, the corresponding
inference $\rho^*({\balpha})$ and $\rho^{*}({\balpha}')$ are also
close to each other. Surprisingly, however, the quantum maximum
entropy inference can be discontinuous! Namely, a small change of
local measurement results may correspond to a dramatic change of the
global quantum state.

The main focus of this work is to relate the discontinuity of the
quantum maximum entropy inference to quantum phase transitions. We
show that the discontinuity of maximum entropy inference signals
level-crossings of the non-local type. That is, at the level-crossing
point, a smooth change of the local Hamiltonian
$H(\boldsymbol{\lambda})$ corresponds to smooth change of the local
density matrices of the ground states, while the change of $\rho^*$,
the maximal entropy inference of these local density matrices is
discontinuous.

We then move on to discuss the possibility of signaling quantum phase
transitions by computing the discontinuity of the maximal entropy
inference $\rho^*$. Given the observation on the relation between
discontinuity of $\rho^*$ and the non-local level-crossings, it is
natural to consider signaling quantum phase transitions by directly
computing where the discontinuity happens. This approach works well in
finite systems, but may fail in the thermodynamic limit of infinite
size systems as the places of discontinuity (i.e. where the system
`closes gap') may change when the system size goes to infinity. Hence,
computations in finite systems may provide no information of the phase
transition point. We propose to solve the problem by using the quantum
conditional mutual information of two disconnected parts of the system
for the ground states. This idea comes from the relationship between
the $3$-body irreducible correlation and quantum conditional mutual
information of gapped systems. As it turns out, the quantum mutual
information works magically well to signature the discontinuity point,
thereby also signals quantum phase transitions in the thermodynamic
limit. In some sense, the quantum conditional mutual information is an
analog of the Levin-Wen topological entanglement
entropy~\cite{levin2006detecting}.

We apply the concept of discontinuity of the maximum entropy inference
to some well-known quantum phase transitions. In particular, we show
that the non-local transition in the ground states of the transverse
quantum Ising chain can be detected by the quantum mutual information
of two disconnect parts of the system. The scope of the applicability
of the quantum conditional mutual information was extended to many
other systems, featuring different types of
transitions~\cite{LIT,zeng2014topological}. All these studies conclude
that the quantum mutual information serves well as a \textit{universal
  indicator} of non-trivial phase transitions.

We organize our paper as follows. In Sec.~\ref{sec:maxent}, we discuss
the concept of the maximum entropy inference and summarize some
important relevant facts. In Sec.~\ref{sec:examples}, we analyze
several examples of discontinuity of the maximum entropy inference
$\rho^*$, ranging from simple examples in dimension $3$ to more
physically motivated ones. In Sec.~\ref{sec:qcmi}, we link the
discontinuity of $\rho^*$ to the concept of the long-range irreducible
many-body correlation and propose to detect the non-local type of
quantum phase transitions by the quantum conditional mutual
information of two disconnect parts of the system. In
Sec.~\ref{sec:further}, further properties of discontinuity of the
maximum entropy inference are discussed. We provide both a necessary
condition and a sufficient condition for the discontinuity to happen.
Finally, Sec.~\ref{sec:summary} contains a summary of all the main
concepts discussed and a discussion of possible future directions.

\section{The Maximum Entropy Inference}
\label{sec:maxent}

We start our discussion by introducing the concept of the maximum
entropy inference given a set of linear constraints on the state
space.

\subsection{The General Case}

Let $\H$ be the $d$-dimensional Hilbert space corresponding to the
quantum system under discussion and $\rho$ be the state of the system.
Let $\D$ be the set of all possible quantum states on $\H$. Any tuple
$\F=(F_1,F_2, \ldots, F_r)$ of $r$ observables defines a mapping
\begin{equation}
  \label{eq:projection}
  \rho \mapsto \balpha = \bigl (\tr(\rho F_1), \tr(F_2 \rho),
  \ldots, \tr(\rho F_r) \bigr),
\end{equation}
from states $\rho$ in $\D$ to points $\balpha$ in the set
\begin{equation*}
  \D_\F = \bigl\{
  \balpha\mid\balpha = \bigl( \tr(\rho
  F_1), \ldots, \tr(\rho F_r) \bigr) \text{ for some } \rho
  \bigr\}.
\end{equation*}
The set $\D_\F$ can be considered as a projection of $\D$ and is a
compact convex set in $\R^r$. If all the $F_i$'s are commuting (i.e.
$[F_i, F_j]=0$, corresponding to the classical case), then $\D_\F$ is
a polytope in $\R^r$.

The convex set $\D_\F$ is mathematically related to the so-called
``(joint) numerical range'' of the operators $F_i$'s. For more
mathematical aspects of these joint numeral ranges and the
discontinuity of the maximum entropy inference, we refer
to~\cite{rodman2015continuity}. We remark that $\D_\F$ is also known
as \textit{quantum convex support} in the
literature~\cite{weis2011quantum}.

As it will be clear in later discussions, the observables $F_i$'s
usually come from the terms in the local Hamiltonian of interest so
that the Hamiltonian is in the span of the observables $F_i$'s. We
will call $H = \sum_i \theta_i F_i$ the Hamiltonian related to the
observables in $\F$. The energy $\tr(H \rho)$ can be written as
\begin{equation*}
  \sum_i \theta_i \tr(\rho F_i),
\end{equation*}
the inner product of the vector $\boldsymbol{\theta} = (\theta_i)$ and
$\balpha$. This means that one can think of the Hamiltonian $H$
geometrically as the supporting hyperplanes of the convex set $\D_\F$.

Given any measurement result $\balpha \in \D_\F$, we are interested in
the set of all states in $\D$ that can give $\balpha$ as the
measurement results. We denote such a set as
\begin{equation*}
  \L(\balpha) = \bigl\{
  \rho \,|\, \tr(\rho F_i)=\alpha_i,\ i=1,\ldots,r
  \bigr\}.
\end{equation*}
It is the preimage of $\balpha$ under the mapping in
Eq.~\eqref{eq:projection}. In other words, it consists of the states
satisfying a set of linear constraints and we call this subset of $\D$
a {\it linear family\/} of quantum states.

In general, there will be many quantum states compatible with
$\balpha$ and $\L(\balpha)$ contains more than one state, unless one
chooses to measure an informationally complete set of observables (for
example, a basis of operators on $\H$ as one often does for the case
of quantum tomography). Especially, when the dimension of system $d$
is large, it is unlikely that one can really measure an
informationally complete set of observables. For instance, for an
$n$-qubit system when $n$ is large, we usually only have access to the
expectation values of local measurements, each involving measurements
only on a few number of qubits. In this case, quantum states
compatible with the local observation data $\balpha$ are usually not
unique.

The question is then what would be the best inference of the quantum
states compatible with the given measurement results $\balpha$. The
answer to this question is well-known, and is given by the principle
of maximum entropy~\cite{Jaynes:1957fy,Wichmann:2004gd}. That is, for
any given measurement results $\balpha$, there is a unique state
$\rho^*\in\L(\balpha)$, given by
\begin{equation}
  \rho^*(\balpha) = \argmax_{\rho \in \L(\balpha)} S(\rho),
\end{equation}
where $S(\rho)$ is the von Neumann entropy of $\rho$. We call
$\rho^*(\balpha)$ the {\it maximum entropy inference\/} for the given
measurement results $\balpha$. More explicitly, it is the optimal
solution of the following optimization problem
\begin{align*}
  \text{Maximize: } & S(\rho)\\
  \text{Subject to: } & \tr(\rho F_i) = \alpha_i,
                        \text{ for all }i=1,2,\ldots,k,\\
                    & \rho \in \D.
\end{align*}

It may seem counter-intuitive that both the maximum entropy inference
$\rho^*$ and its entropy can be discontinuous~\cite{knauf2010entropy}
as functions of the local measurement data $\balpha$. When we say
$\rho^*$ is discontinuous, we mean the state itself, not its entropy,
is discontinuous. Indeed there could be examples where these two
concepts are not the same (e.g. the energy gap of the system closes
but the ground-state degeneracy does not change). For all examples
considered in this paper, however, the entropy is also discontinuous
when the state is. We note that the discontinuity of the maximum
entropy inference is a genuinely quantum effect as the classical
maximum entropy inference is always
continuous~\cite{Jaynes:1957fy,Wichmann:2004gd}.

\subsection{The Case of Local Measurements}

The discussions in the above subsection specialize to the important
case of many-body physics with local measurements.

Consider an $n$-particle system where each particle has dimension $d$.
The Hilbert space $\H$ of the systems is $(\C^{d})^{\otimes n}$, with
dimension $d^n$. We know that, for an $n$-particle state $\rho$, we
usually only have access to the measurement results of a set of local
measurements $\F=(F_1,\ldots, F_r)$ on the system, where each $F_i$
acts on at most $k$ particles for $k\leq n$. The most interesting case
is where $n$ is large and $k$ is small (usually a constant independent
of $n$). In this sense, we will just call such a measurement setting
$k$-local.

Notice that each measurement result $\tr(\rho F_i)$ now depends only
on the $k$-particle reduced density matrix ($k$-RDM) of the particles
that $F_i$ is acting non-trivially on. It is convenient to write the
set of all the $k$-RDMs of $\rho$ (in some fixed order) as a vector
$\brho^{(k)}=\{\rho^{(k)}_1,\ldots,\rho^{(k)}_m\}$, where each
component is a $k$-RDM of $\rho$ and $m={n\choose k}$. The $k$-RDMs
$\brho^{(k)}$ will play the role of expectation values
$\boldsymbol{\alpha}$ as in the general case.

Along this line, the set of results of all $k$-local measurements can
be defined in terms of $k$-RDMs, and we write the set $\D^{(k)}$ of
all such measurement results as
\begin{equation}
  \D^{(k)} = \bigl\{
  \brho^{(k)} \,\mid\, \brho^{(k)}
  \text{ is the } k\text{-RDMs of some } \rho
  \bigr\}.
\end{equation}
Similarly, the linear family can also be defined in terms of $k$-RDMs,
\begin{equation}
  \L(\brho^{(k)}) = \bigl\{ \rho \,\mid\, \rho
  \text{ has the }k\text{-RDMs}\ \brho^{(k)} \bigr\}.
\end{equation}
The maximum entropy inference given the $k$-RDMs $\brho^{(k)}$ is
\begin{equation}
  \rho^{*}(\brho^{(k)}) = \argmax_{\rho \in\L(\brho^{(k)})} S(\rho).
\end{equation}

We remark that, in practice, one may not be interested in all the
$m={n\choose k}$ $k$-RDMs, but rather only those $k$-RDMs that are
{\it geometrically\/} local. For instance, for a lattice spin model,
one may only be interested in the $2$-RDMs of the nearest-neighbour
spins. Our discussion can also be generalized to these cases, as in
the discussion in~\cite{zeng2014topological} for one-dimensional spin
chains. There could also be cases that the system has certain symmetry
(for instance a bosonic system or fermionic system where all the
$k$-RDMs are the same), and our theory can be naturally adapted to
these cases.

The maximum entropy inference $\rho^*$ given local density matrices
has a more concrete physical meaning. For any $n$-particle state
$\rho$, if $\rho=\rho^*(\brho^{(k)})$, then $\rho$ is uniquely
determined by its $k$-RDMs using the maximum entropy principle. One
can argue, in this case, that all the information (including all
correlations among particles) contained in $\rho$ are already
contained in its $k$-RDMs. In other words, $\rho$ does not contain any
irreducible correlation~\cite{zhou2008irreducible} of order higher
than $k$. On the other hand, if $\rho\neq\rho^*$, then $\rho$ cannot
be determined by its $k$-RDMs and there are more
information/correlations in $\rho$ than those in its $k$-RDMs.
Therefore, $\rho$ contains non-local irreducible correlation that
cannot be obtained from its local RDMs.

\section{Discontinuity of $\rho^*$}
\label{sec:examples}
In this section, we explore the discontinuity of $\rho^*$ based on
several simple examples. The first three of them involve only two
different measurement observables, but they do demonstrate almost all
the key ideas in the general case.

\subsection{The Examples of Two Observables}

We will choose $d=3$ for the Hilbert space dimension as it is enough
to demonstrate most of the phenomena we need to see. Fix an arbitrary
orthonormal basis of $\C^3$, say, $\{ \ket{0}, \ket{1}, \ket{2}\}$.

\begin{example}
  \label{eg:numrange1}
  $\F$ consists of the following two observables
  \begin{equation}
    F_1 = \begin{pmatrix}
      1 & 0 & 0 \\ 0 & 1 & 0 \\ 0 & 0 & -1
    \end{pmatrix}, %
    \quad %
    F_2 = \begin{pmatrix} %
      1 & 0 & 1 \\ 0 & 1 & 1 \\ 1 & 1 & -1
    \end{pmatrix}.
  \end{equation}
\end{example}

\begin{figure}[ht]%
  \centering %
  \subfloat[][]{%
      \label{fig:numrange1}     %
      \includegraphics[scale=0.35]{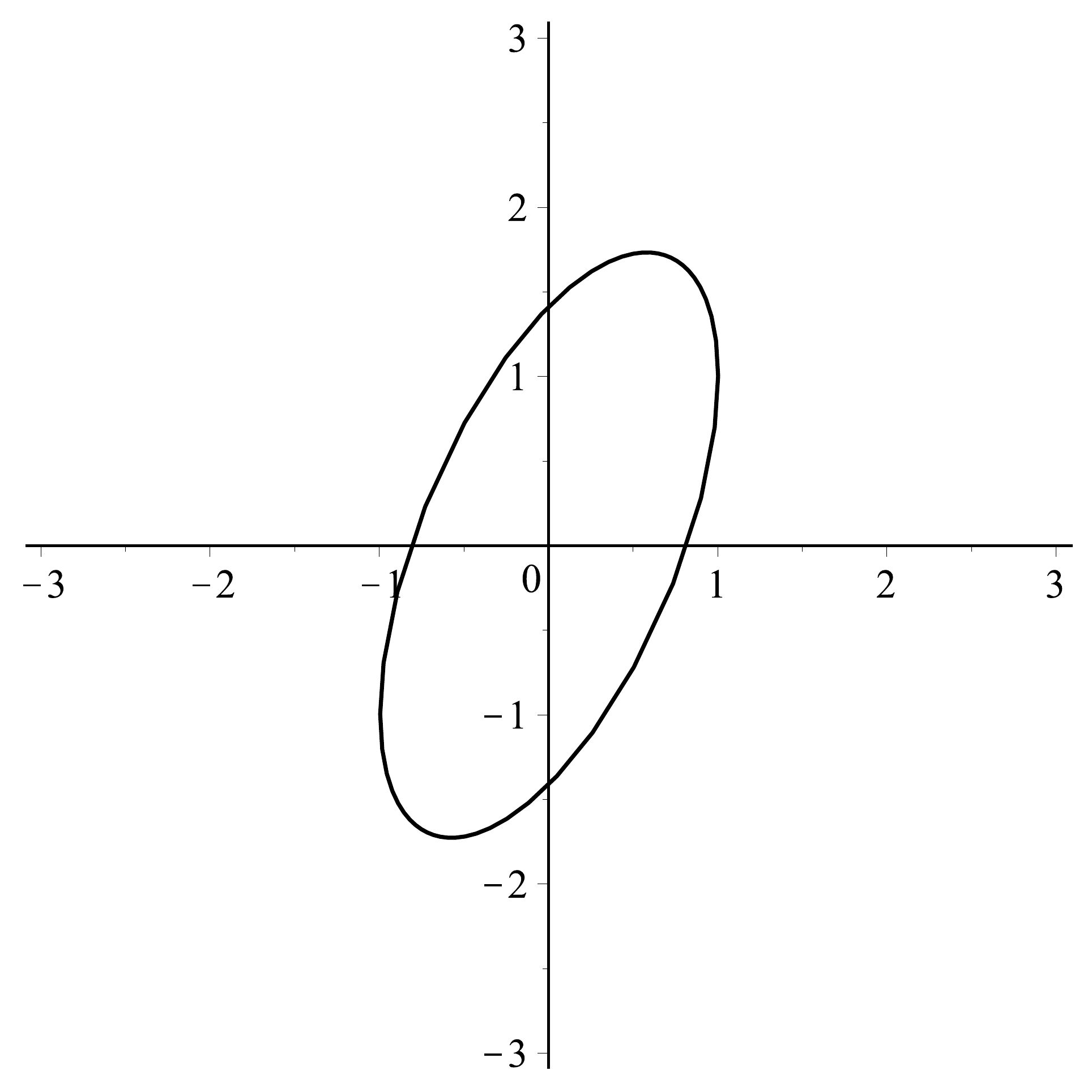}
    }%
    \hspace{8pt}%
    \subfloat[][]{%
      \label{fig:lines1}%
      \includegraphics[scale=0.35]{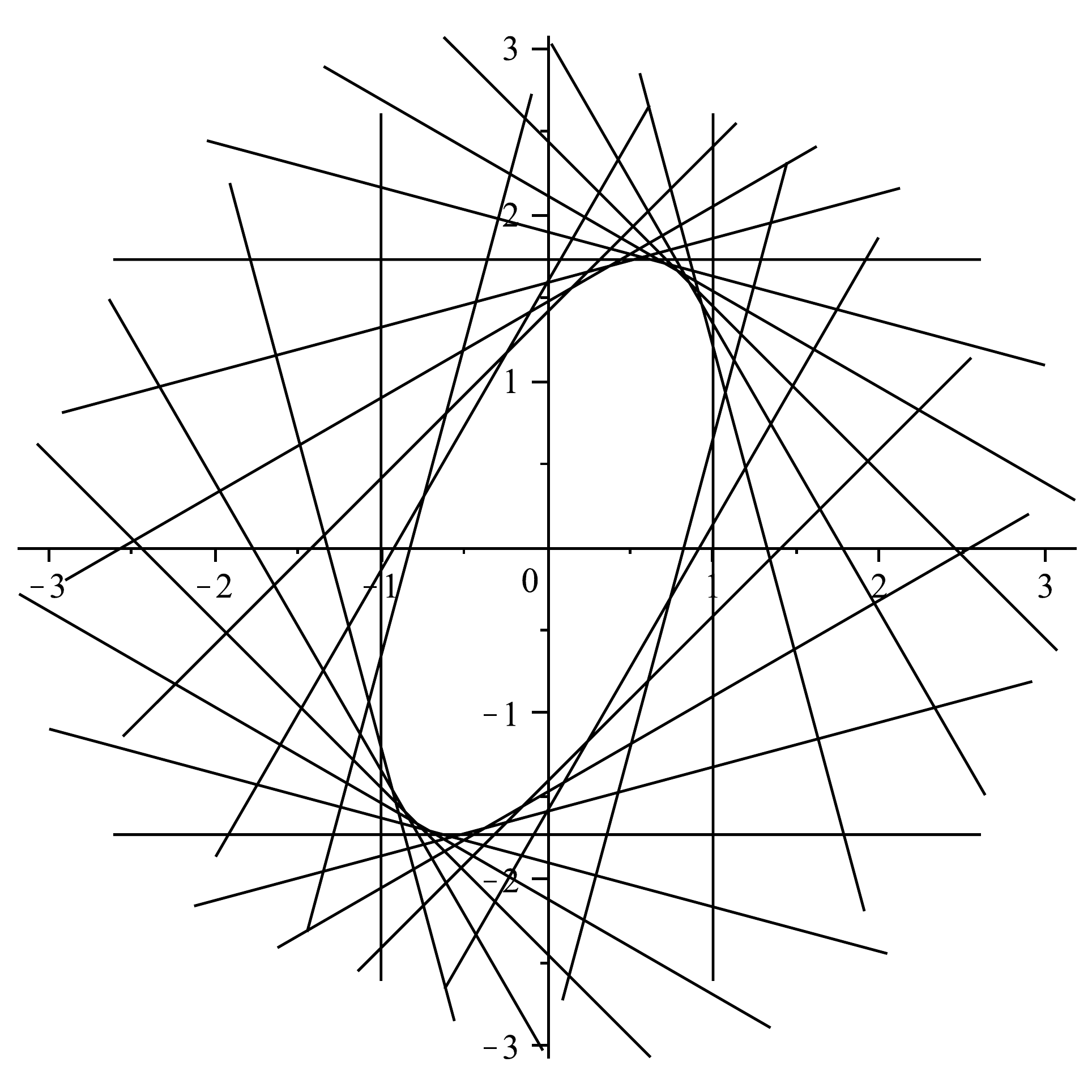}}
    \caption[]{\subref{fig:numrange1} The convex set of $\D_\F$ in
      $\R^2$. The horizontal axis corresponds to the value of
      $\tr(\rho F_1)$ and the vertical axis corresponds to
      $\tr(\rho F_2)$; \subref{fig:lines1} The supporting hyperplanes
      of $\D_\F$ in $\R^2$ (i.e. the straight lines on the figure
      which are tangent to $\D_\F$), which corresponds to the
      Hamiltonians $H=\theta_1 F_1+\theta_2 F_2$.}
    \label{fig:nrl1}%
  \end{figure}

First, notice that $F_1, F_2$ do not commute. The set of all possible
measurement results $\D_\F$ is a convex set in $\R^2$. We plot this
convex set in Fig.~\ref{fig:nrl1}~\subref{fig:numrange1}. To obtain
this figure, we let $\rho$ vary for all the density matrices on
$\C^3$, and let the corresponding $\tr(\rho F_1)$ be the horizontal
coordinate and $\tr(\rho F_2)$ the vertical coordinate. The resulting
picture is nothing but the numerical range of the matrix
$F_1 + i F_2$.

As discussed in Sec.~\ref{sec:maxent}, the Hamiltonian $H$ related to
$\F$ has the form
\begin{equation}
  H = \theta_1 F_1+\theta_2 F_2
\end{equation}
for some parameters $\theta_1, \theta_2 \in \R$. Notice that the
Hamiltonian corresponds to supporting hyperplanes of $\D_\F$, as the
inner product has the form
$\tr(H\rho)=(\theta_1, \theta_2)\cdot(\alpha_1,\alpha_2)^T$. We
demonstrate these supporting hyperplanes of $\D_\F$ in
Fig.~\ref{fig:nrl1}~\subref{fig:lines1}.


It is straightforward to see that the ground state of $H$ is
non-degenerate except for the case $\theta_1<0, \theta_2=0$, where the
ground space is two-fold degenerate with a basis $\{\ket{0},\ket{1}\}$
corresponding to the measurement results $\balpha_0=(1,1)$.

We now show that the maximum entropy inference $\rho^*(\balpha)$ is
indeed discontinuous at the point $\balpha_0=(1,1)$. To see this,
first notice that the corresponding
$\rho^* = \frac{1}{2} (\ket{0} \bra{0} + \ket{1} \bra{1})$. While for
any small $\epsilon$, the corresponding ground state space of
$-F_1 + \epsilon F_2$ is no longer degenerate, which means
$\rho^*(\balpha)$ is a {\it pure\/} state for $\balpha\neq \balpha_0$.

Therefore, for any sequence of $\balpha$ on the boundary of $\D_\F$
approaching $\balpha_0$,
\begin{equation}
  \rho^*(\balpha)\not\rightarrow\rho^*(\balpha_0)
  \ \text{when}\ {\balpha\rightarrow \balpha_0},
\end{equation}
and the discontinuity of $\rho^*(\balpha)$ follows.

This example seems to indicate that the discontinuity simply comes
from degeneracy: as in general degeneracy is rare, whenever such a
point of degeneracy exists, we have a singularity on the boundary of
$\D_\F$ so discontinuity happens. However, it is important to point
out that this is not quite true. For example, degeneracy also happens
in classical systems where there can have no discontinuity of
$\rho^*$. We further explain this point in the following example.

\begin{example}
  \label{eg:numrange2}
  $\F$ consists of the following two observables

  \begin{equation}
    F_1 = \begin{pmatrix}
      1 & 0 & 0 \\ 0 & 1 & 0 \\ 0 & 0 & -1
    \end{pmatrix}, %
    \quad %
    F_2 = \begin{pmatrix} 1 & 0 & 1 \\ 0 & 0 & 1 \\ 1 & 1 & -1
    \end{pmatrix}.
  \end{equation}
\end{example}

\begin{figure}[ht]%
  \centering %
  \subfloat[][]{%
      \label{fig:numrange2}     %
      \includegraphics[scale=0.35]{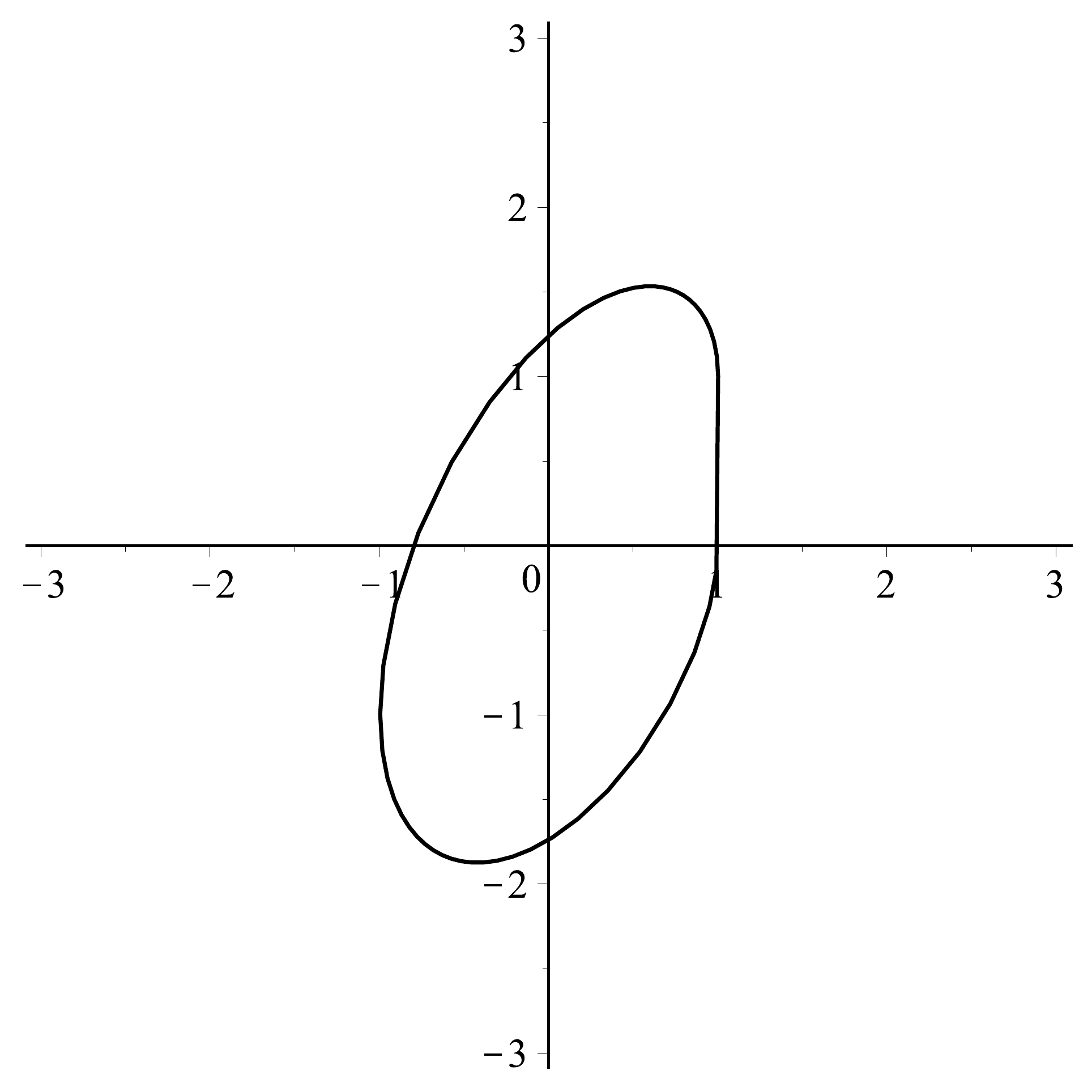}
    }%
    \hspace{8pt}%
    \subfloat[][]{%
      \label{fig:lines2}%
      \includegraphics[scale=0.35]{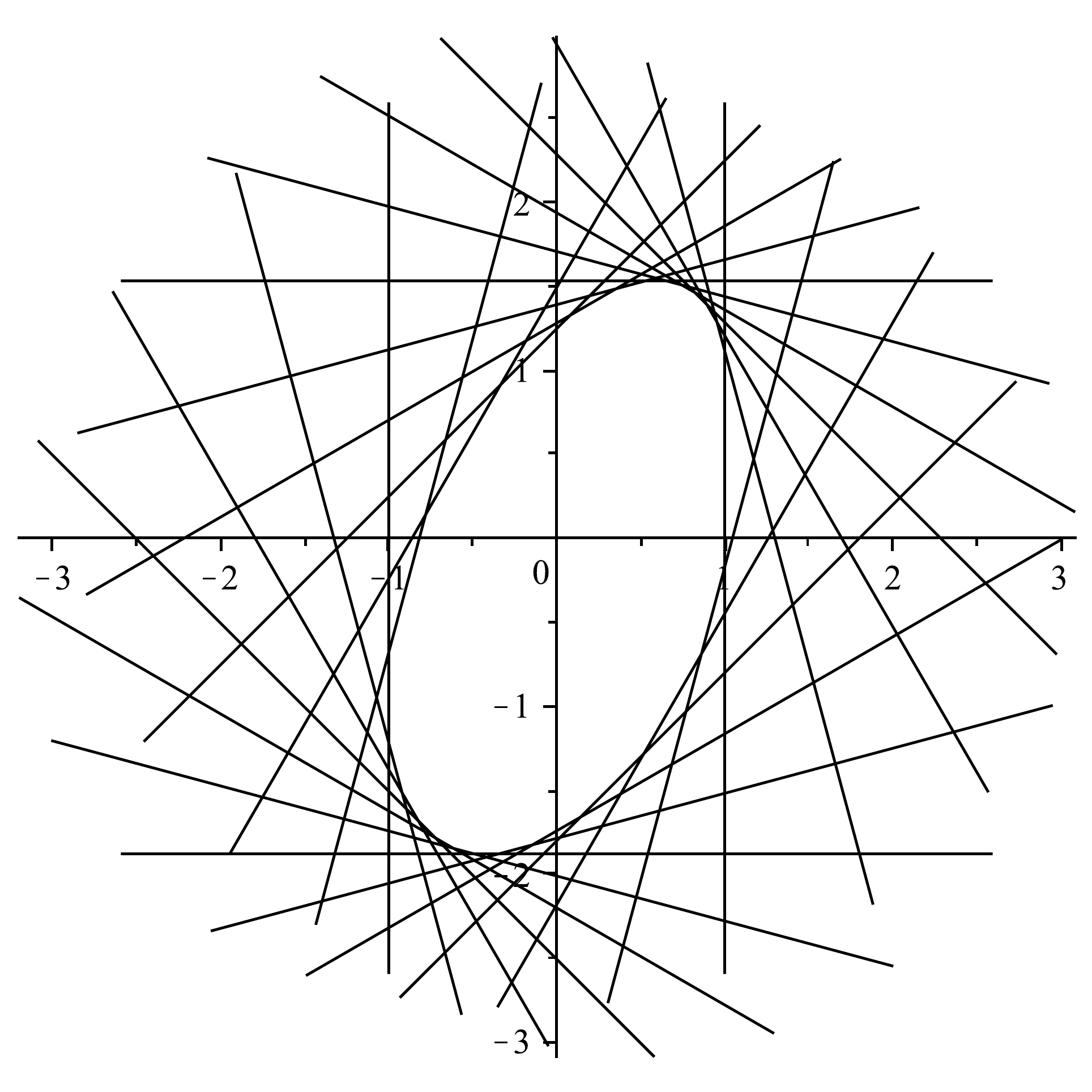}}
    \caption[]{\subref{fig:numrange2} The convex set of $\D_\F$ in
      $\R^2$. The horizontal axis corresponds to the value of
      $\tr(\rho F_1)$ and the vertical axis corresponds to
      $\tr(\rho F_2)$; \subref{fig:lines2} The supporting hyperplanes
      of $\D_\F$ in $\R^2$ (i.e. the straight lines on the figure
      which are tangent to $ \D_\F $), which corresponds to the
      Hamiltonians $H=\theta_1 F_1+\theta_2 F_2$. }
    \label{fig:nrl2}%
  \end{figure}

Notice that again $[F_1,F_2]\neq 0$. And we show the convex set
$\D_\F$ in Fig.~\ref{fig:nrl2}~\subref{fig:numrange2}.

  Consider the Hamiltonian $H=\theta_1 F_1 + \theta_2 F_2$ for some
  $\theta_1, \theta_2 \in \R$, as illustrated as supporting
  hyperplanes in Fig.~\ref{fig:nrl2}~\subref{fig:lines2}. Similarly,
  the ground state of $H$ is two-fold degenerate for
  $\theta_1<0, \theta_2=0$ (corresponding to the vertical line at
  $\alpha_1=1$) with a basis $\{\ket{0},\ket{1}\}$. However, different
  from Example~\ref{eg:numrange1}, the ground states
  do not correspond to a single measurement result $\balpha_1=(1,1)$.
  Instead, they are on the line $[(1,0),(1,1)]$.


By simple calculations, now the maximum entropy inference
$\rho^*(\balpha)$ is in fact continuous at the point
$\balpha_1=(1,1)$, and on the entire line $[(1,0),(1,1)]$. In fact,
$\rho^*(\balpha_p) = p\ket{0}\bra{0} + (1-p)\ket{1}\bra{1}$ for
$\balpha_p = (1,p)$.

For any small perturbation $\epsilon$, the corresponding ground state
space of $-F_1+\epsilon F_2$ is non-degenerate, meaning
$\rho^*(\balpha)$ is a pure state. This change of $\rho^*(\balpha)$
from $\epsilon<0$ to $\epsilon>0$ is {\it sudden\/} with respect to
the small change of $\epsilon$, which, however, is accompanied by a
sudden change also in the measurement results (from a point near
$(1,1)$ to $(1,0)$). As we are considering the discontinuity of
$\rho^*$ with respect to the measurement data $\balpha$, not the
parameter $\epsilon$ in the Hamiltonian, $\rho^*$ is in fact
continuous.

This example demonstrates that when Hamiltonian changes smoothly,
ground states have sudden changes accompanied with the sudden change
of measurement results. In other words, the change of ground states
can be described already by the change of local measurement results.
This is somewhat a classical feature, as discussed in the next
example.

\begin{example}
  \label{eg:numrange3}
  $\F$ consists of the following two observables
  \begin{equation}
    F_1 = \begin{pmatrix}
      1 & 0 & 0 \\ 0 & 1 & 0 \\ 0 & 0 & -1
    \end{pmatrix}, %
    \quad %
    F_2=\begin{pmatrix} 1 & 0 & 0 \\ 0 & 0 & 0 \\ 0 & 0 & -1
    \end{pmatrix}.
  \end{equation}
\end{example}

Now this corresponds to the classical situation where $[F_1,F_2]=0$.

\begin{figure}[ht]%
  \centering %
  \subfloat[][]{%
      \label{fig:numrange3}     %
      \includegraphics[scale=0.35]{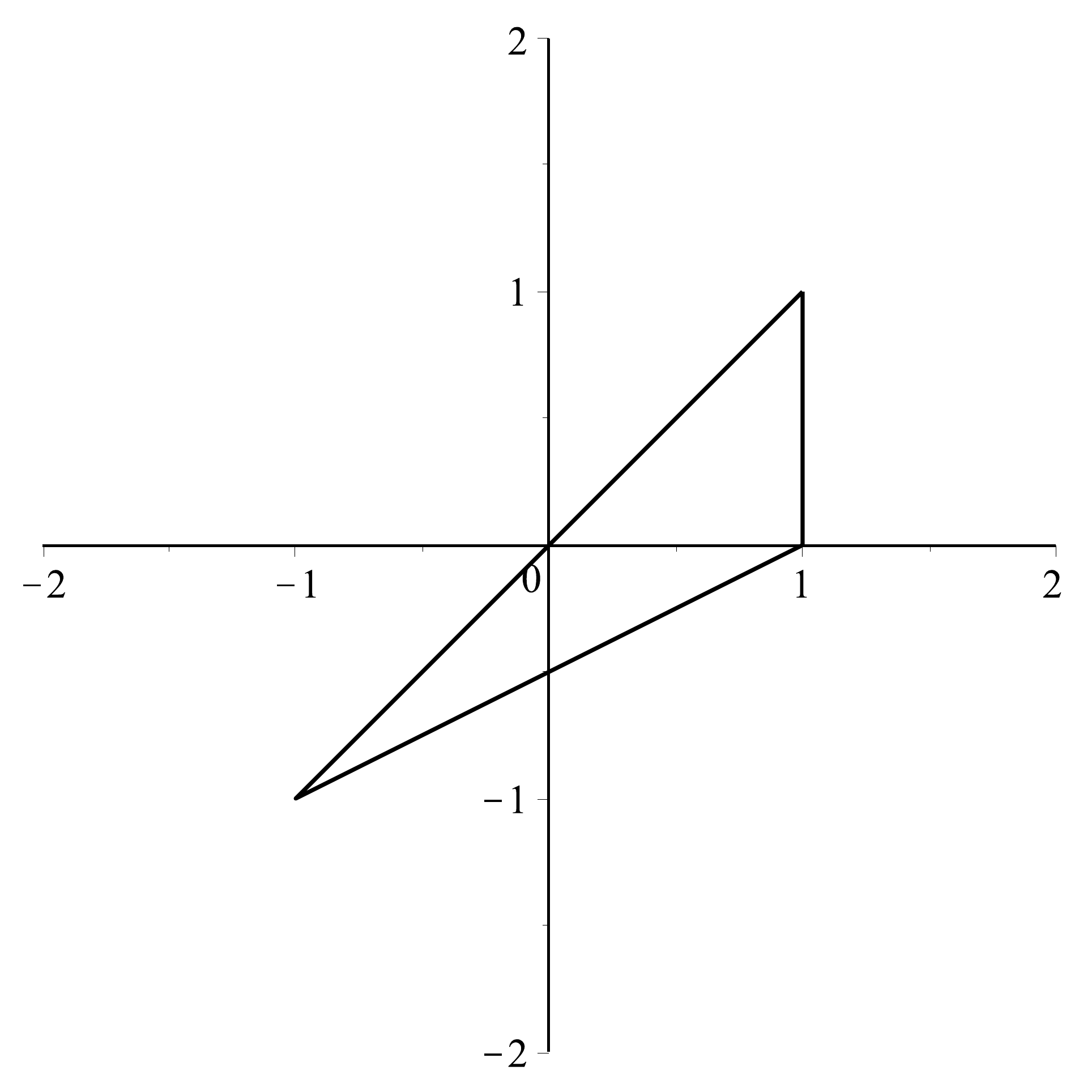}
    }%
    \hspace{8pt}%
    \subfloat[][]{%
      \label{fig:lines3}%
      \includegraphics[scale=0.35]{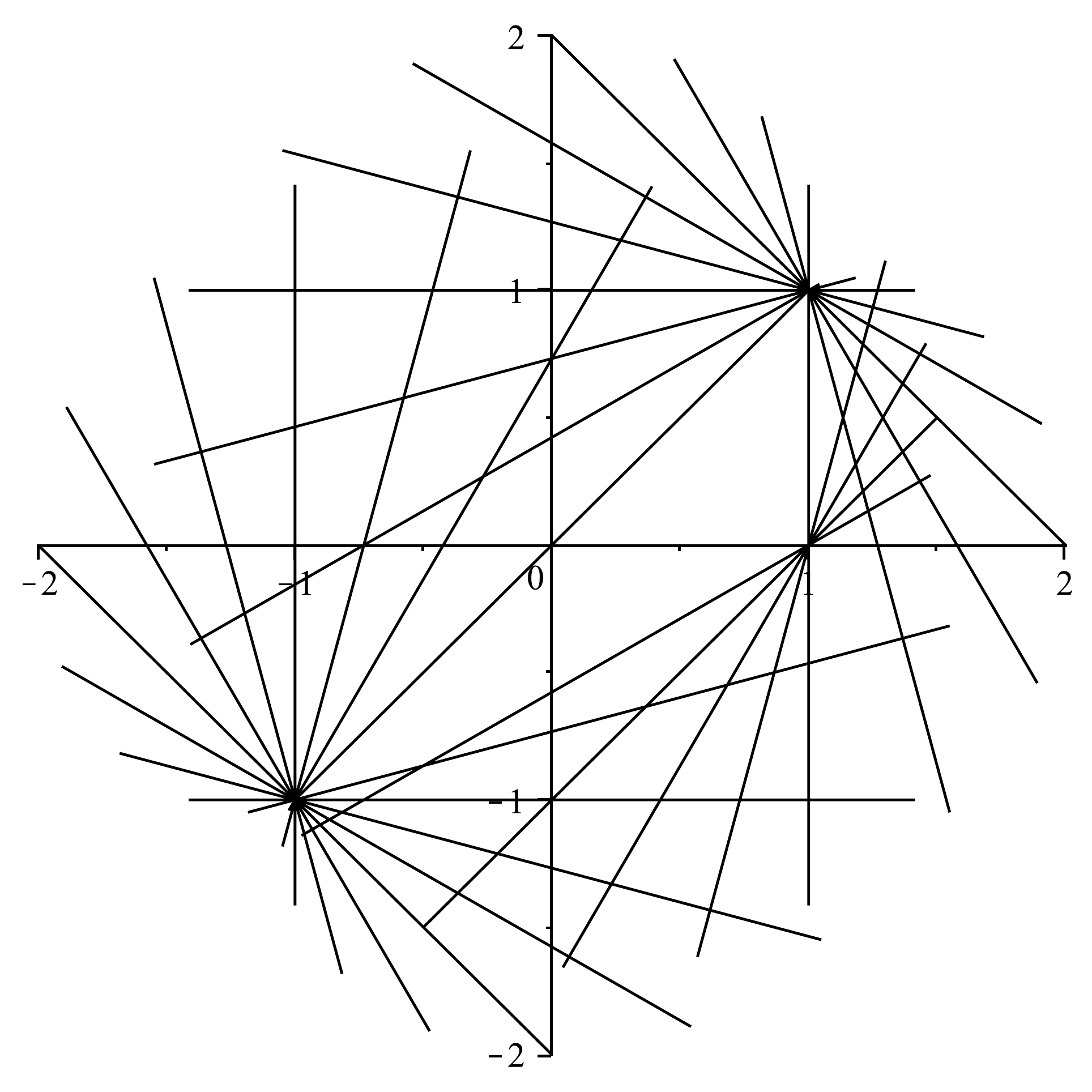}
    }
    \caption[]{\subref{fig:numrange3} The convex set of $\D_\F$ in
      $\R^2$. The horizontal axis corresponds to the value of
      $\tr(\rho F_1)$ and the vertical axis corresponds to
      $\tr(\rho F_2)$; \subref{fig:lines3} The supporting hyperplanes
      of $\D_\F$ in $\R^2$ (i.e. the straight lines on the figure
      which are tangent to $\D_\F$), which corresponds to the
      Hamiltonians $H=\theta_1 F_1+\theta_2 F_2$.}
    \label{fig:nrl3}%
  \end{figure}

The convex set $\D_\F$ in given in
Fig.~\ref{fig:nrl3}~\subref{fig:numrange3}. It is a triangle for this
example, and a polytope in the general classical case.

  Consider the related Hamiltonian $H=\theta_1 F_1 + \theta_2 F_2$ for
  some $\theta_1, \theta_2 \in \R$, as illustrated as supporting
  hyperplanes in Fig.~\ref{fig:nrl3}~\subref{fig:lines3}. Similarly,
  the ground state of $H$ is two-fold degenerate for
  $\theta_1<0, \theta_2=0$ (corresponding to the vertical line at
  $\alpha_1=1$) with a basis $\{\ket{0},\ket{1}\}$. For a similar
  reason, the maximum entropy inference $\rho^*(\balpha)$ is
  continuous on the entire line $[(1,0),(1,1)]$ as in the previous
  example.


If we still consider for any small perturbation $-F_1+\epsilon F_2$,
the corresponding ground-state space is non-degenerate: it is
$\ket{1}$ for $\epsilon<0$ and $\ket{2}$ for $\epsilon>0$. So from
$\epsilon<0$ to $\epsilon>0$, we also see sudden changes of both the
measurement results and the ground states.




In the above three examples, the first one is the most interesting and
exhibits smooth change in measurement results and discontinuity of the
maximum entropy inference $\rho^*$. The second and third behave in a
similar classical way where a small change in the Hamiltonian will
induce a sudden change of measurement results and there is no
discontinuity of $\rho^*$. We summarize our observations from the
three examples in this subsection as below. Although the examples
involve two observables only, we state the observation in the more
general setting of arbitrarily many observables.

\begin{observation}
  \label{obs:type}
  Given a set of measurements $\F = (F_1, F_2, \ldots, F_r)$, and a
  family of related Hamiltonians $H$ of the form
  $H=\sum_i \theta_i F_i$ with $\theta_i$ changing with certain
  parameter. The Hamiltonian $H$ has two types of ground state level
  crossing:
  \begin{itemize}
  \item Type I (local type): level-crossing that can be detected by a
    sudden change of the measurement results.
  \item Type II (non-local type): level-crossing that cannot be
    detected by a sudden change of the measurement results.
  \end{itemize}
  More importantly, only Type II corresponds to discontinuity of the
  maximum inference $\rho^*(\balpha)$.
\end{observation}

\subsection{The Example of Local Measurements}

We now give a simple example showing the discontinuity of $\rho^*$ in
a three-qubit system with $2$-local interactions.

\begin{example}
  \label{eg:GHZ}
  The three-qubit GHZ state given by
  \begin{equation}
    \ket{\text{GHZ}_3}=\frac{1}{\sqrt{2}}(\ket{000}+\ket{111})
  \end{equation}
  is known to be the ground state of a two-body Hamiltonian
  \begin{equation}
    H=-Z_1Z_2-Z_2Z_3
  \end{equation}
  with $Z_i$ the Pauli $Z$ operator acting on the $i$-th qubit. The
  ground-state space of $H$ is two-fold degenerate and is spanned by
  $\{\ket{000},\ket{111}\}$. Now consider the $2$-RDMs of the GHZ state
  \begin{equation}
    \brho^{(2)}=\{\rho_{\{1,2\}},\rho_{\{2,3\}},\rho_{\{1,3\}}\},
  \end{equation}
  with $\rho_{\{i,j\}}=\frac{1}{2}(\ket{00}\bra{00}+\ket{11}\bra{11})$
  being the $2$-RDM of qubits $i$ and $j$. We claim that there is
  discontinuity at $\brho^{(2)}$.
\end{example}

To see this, consider a family of perturbations
$H + \epsilon \sum_{i=1}^3 X_i$ of the Hamiltonian $H$. For any
$\epsilon \ne 0$, the ground space is non-degenerate and the unique
ground state converges to $\ket{\text{GHZ}_3}$ when
$\epsilon \rightarrow 0-$ and to $(\ket{000}-\ket{111})/\sqrt{2}$ when
$\epsilon \rightarrow 0+$. As the ground state is unique when
$\epsilon \ne 0$ and the Hamiltonian is $2$-local, the $2$-RDMs of the
ground state determines the state. This means that $\rho^*$ is pure
and coincide with the ground state for all $\epsilon \ne 0$. However,
at $\epsilon = 0$, $\rho^*$ is
\begin{equation}
  \rho^*(\brho^{(2)}) = \frac{1}{2}(\ket{000} \bra{000}
  + \ket{111}\bra{111}),
\end{equation}
and the discontinuity of $\rho^*$ follows.

It is worth pointing out the similarity in the structure of the above
example and Example~\ref{eg:numrange1}, despite their totally
different specific form. First notice that
$\frac{1}{\sqrt{2}}\ket{000}\pm\ket{111}$ are two eigenstates of
$Z_1Z_2+Z_2Z_3$ of the same eigenvalue $1$. If we complete
$\frac{1}{\sqrt{2}}\ket{000}\pm\ket{111}$ to a basis, $Z_1Z_2+Z_2Z_3$
will have a $2$-by-$2$ identity block with zero entries to the right
and bottom. In that basis, the $\sum_{i=1}^3 X_i$ also has such a
$2$-by-$2$ block proportional to identity and has some non-zero off
diagonal entries. In other words, $Z_1Z_2+Z_2Z_3$ and
$\sum_{i=1}^3 X_i$ has a rather similar block structure as $F_1$ and
$F_2$ in Example~\ref{eg:numrange1}.

We generalize the Observation~\ref{obs:type} in terms of local
measurements as follows.

\begin{obsprimed}{obs:type}
  \label{obs:typeb}
  For an $n$-particle system, consider the set of all $k$-local
  measurements $\F$, which then corresponds to a local Hamiltonian
  $H=\sum_j c_jF_j$ with $F_j\in\F$ acting nontrivially on at most $k$
  particles. There are two kinds of ground state level crossing:
  \begin{itemize}
  \item Type I: level-crossing that can be detected by a sudden change
    of the $k$-RDMs $\brho^{(k)}$.
  \item Type II: level-crossing that cannot be detected by a sudden
    change of the local $k$-RDMs $\brho^{(k)}$.
  \end{itemize}
  Only Type II corresponds to discontinuity of the maximum entropy
  inference $\rho^*(\brho^{(k)})$.
\end{obsprimed}

\subsection{The Example of Transverse Quantum Ising Model}
\label{sec:tqic}

Our next example is an $n$-qubit generalization of
Example~\ref{eg:GHZ} and is known as the transverse quantum Ising
model.

\begin{example}
  \label{eg:Ising}
  The Ising Hamiltonian is given by
  \begin{equation}
    H(\lambda)=-J(\sum_{i=1}^{n-1} Z_iZ_{i+1}+\lambda \sum_{i=1}^n X_i),
  \end{equation}
  for $J>0$. For any finite $n$ the discontinuity of $\rho^*$
  determined by the $2$-RDMs happen at $\lambda = 0$. For infinite $n$,
  the discontinuity of $\rho^*$ happen at $\lambda = 1$.
\end{example}

The Hamiltonian $H(\lambda)$ has a $\mathbb{Z}_2$ symmetry, which is
given by $X^{\otimes n}$, i.e. $[X^{\otimes n},H(\lambda)]=0$. In the
limit of $\lambda=0$, the ground state of $H(0)$ is two-fold
degenerate and spanned by
$\{ \ket{0}^{\otimes n}, \ket{1}^{\otimes n} \}$. And in the limit of
$\lambda=\infty$, the ground state of $H(\infty)$ is non-degenerate
and is given by $\frac{1}{\sqrt{2}} (\ket{0} + \ket{1})^{\otimes n}$.

In the case of finite $n$, the ground space of $H(\lambda)$ for any
$\lambda>0$ is non-degenerate. Based on a similar discussion of
Example~\ref{eg:GHZ}, we have
\begin{equation}
  \lim_{\lambda\rightarrow 0^{+}} \rho^*(\lambda) = \ket{\text{GHZ}_n}
  \bra{\text{GHZ}_n},
\end{equation}
where $\ket{\text{GHZ}_n}$ is the $n$-qubit GHZ state
$\frac{1}{\sqrt{2}}(\ket{0}^{\otimes n}+\ket{1}^{\otimes n})$. On the
other hand, at $\lambda=0$, $\rho^*(0)$ has rank $2$. When the
$2$-RDMs of $\rho^*(0)$ is approached by the $2$-RDMs of the ground
states $\rho^*(\lambda)$ of $H(\lambda)$, the local RDMs of
$\rho^*(\lambda)$ change smoothly, and discontinuity of
$\rho^*(\lambda)$ happens at $\lambda=0$.

In the thermodynamic limit of $n\rightarrow\infty$, it is well-known
that when $\lambda$ increases from $0$ to $\infty$, quantum phase
transition happens at the point $\lambda=1$~\cite{pfeuty1970one}. For
$\lambda\rightarrow 1^{+}$, $\lambda=1$ is exactly the point where the
ground space of $H(\lambda)$ undergoes the transition from
non-degenerate to degenerate. A discontinuity of $\rho^{*}(\lambda)$
happens at $\lambda=1$ when $\lambda\rightarrow 1^{+}$, which is a
sudden jump of rank from $1$ to $2$, while the local RDMs of
$\rho^{*}(\lambda)$ change smoothly.

For $0<\lambda\leq 1$, the two-fold degenerate ground states, although
not exactly the same as those two at $\lambda=0$, are qualitatively
similar. For the range of $0\leq\lambda\leq 1$, the ground states are
all two-fold degenerate. For finite $n$, however, in the region of
$0<\lambda\leq 1$, an (exponentially) small gap exists between two
near degenerate states, and the true ground state does not break the
$\mathbb{Z}_2$ symmetry of the Hamiltonian $H(\lambda)$.

This example demonstrates the dramatic difference between the case of
finite $n$ and the case of the thermodynamic limit of infinite $n$. It
also foretells the difficulty of signaling phase transitions by
computing the discontinuity of $\rho^*$ of finite systems directly. We
will propose a solution to this problem in Sec.~\ref{sec:qcmi}.

\section{Signaling Discontinuity by Quantum Conditional Mutual
  Information}
\label{sec:qcmi}
\subsection{Irreducible Correlation and Quantum Conditional Mutual
  Information}
\label{sec:corr}

We have mentioned the relation between the maximum entropy inference
and the theory of irreducible many-body
correlations~\cite{zhou2008irreducible}. For an $n$-particle quantum
state $\rho$, denote its $k$-RDMs by $\brho^{(k)}$. Then its
$k$-particle irreducible correlation is given
by~\cite{linden2002almost,zhou2008irreducible}
\begin{equation}
  C^{(k)}(\rho) = S(\rho^*(\brho^{(k-1)})) -
  S(\rho^*(\brho^{(k)})).
\end{equation}
What $C^{(k)}$ measures is the amount of correlation contained in
$\brho^{(k)}$ but not contained in $\brho^{(k-1)}$.

Consider a partition $A,B,C$ of the $n$ particles so that $A$ and $C$
are far apart. Define
\begin{equation}
  \label{eq:rhoABC}
  \rho^*_{ABC} = \argmax_{\displaystyle \sigma_{AB}=\rho_{AB} \atop
    \displaystyle \sigma_{BC}=\rho_{BC} } S(\sigma_{ABC}).
\end{equation}
Then the three-body irreducible correlation of $\rho_{ABC}$ is given
by
\begin{equation}
  C^{ABC} = S(\rho^*_{ABC}) - S(\rho_{ABC}).
\end{equation}
Note that we do not include the constraint $\sigma_{AC} = \rho_{AC}$
in the definition of $\rho^*_{ABC}$. The reason for this is that the
region of $A$ and $C$ are chosen to be far apart and, therefore, there
will be no $k$-local terms in the Hamiltonian that act non-trivially
on both $A$ and $C$.

In the discussion on the example of quantum Ising chain, we have
observed the difficulty of signaling the discontinuity of $\rho^*$ in
the thermodynamic limit by computations of finite systems. In the
following, we propose a quantity that can reveal the physics in the
thermodynamic limit by investigating relatively small finite systems.

The quantity we will use is the quantum conditional mutual information
\begin{equation}
  I(A\,{:}\,C|B)_\rho=S(\rho_{AB})+S(\rho_{BC})-S(\rho_B)-S(\rho_{ABC}).
\end{equation}
We will also omit the subscript $\rho$ when there is no ambiguity.
Usually, the state $\rho$ will be chosen to be a reduced state of the
ground state of the Hamiltonian. It is known that the quantum
conditional mutual information is an upper bound of
$C^{ABC}$~\cite{levin2006detecting,2014arXiv1402.4245L}. Namely, we
have
\begin{equation}
  \label{eq:CIACB}
  C^{ABC}(\rho) \leq I(A\,{:}\,C|B)_\rho,
\end{equation}
which is equivalent to the strong subadditivity~\cite{LR73} for the
state $\rho^*_{ABC}$. The equality holds when the state $\rho^*_{ABC}$
satisfies $I(A\,{:}\,C|B) = 0$, or is a so-called quantum Markovian
state.

We will use the quantum conditional mutual information
$I(A\,{:}\,C|B)$ of the ground state, instead of $3$-body irreducible
correlation $C^{ABC}$, to signal the discontinuity and phase
transitions in the system. We do this for two reasons. First, it is
conjectured that the equality in Eq.~\eqref{eq:CIACB} always holds in
the thermodynamic limit for gapped systems. In other words, the
corresponding $\rho^*_{ABC}$ of the ground state is always a quantum
Markovian state (there are reasons to believe this, see
e.g.~\cite{2014arXiv1402.4245L,kim2014informational}). Assuming this
conjecture, $I(A\,{:}\,C|B)$ is indeed a good quantity to signal the
discontinuity and phase transition in the thermodynamic limit. Second,
as it turns out, quantum conditional mutual information performs much
better as in indicator when we do computations in systems of small
system sizes. Most importantly, it doesn't seem to suffer from the
problem $C^{ABC}$ has in finite systems. For more discussion on the
physical aspects of $I(A\,{:}\,C|B)$, we refer to~\cite{LIT}.

\subsection{The Transverse Ising Model}

\begin{figure}[ht]
  \centering
  \includegraphics[scale=0.25]{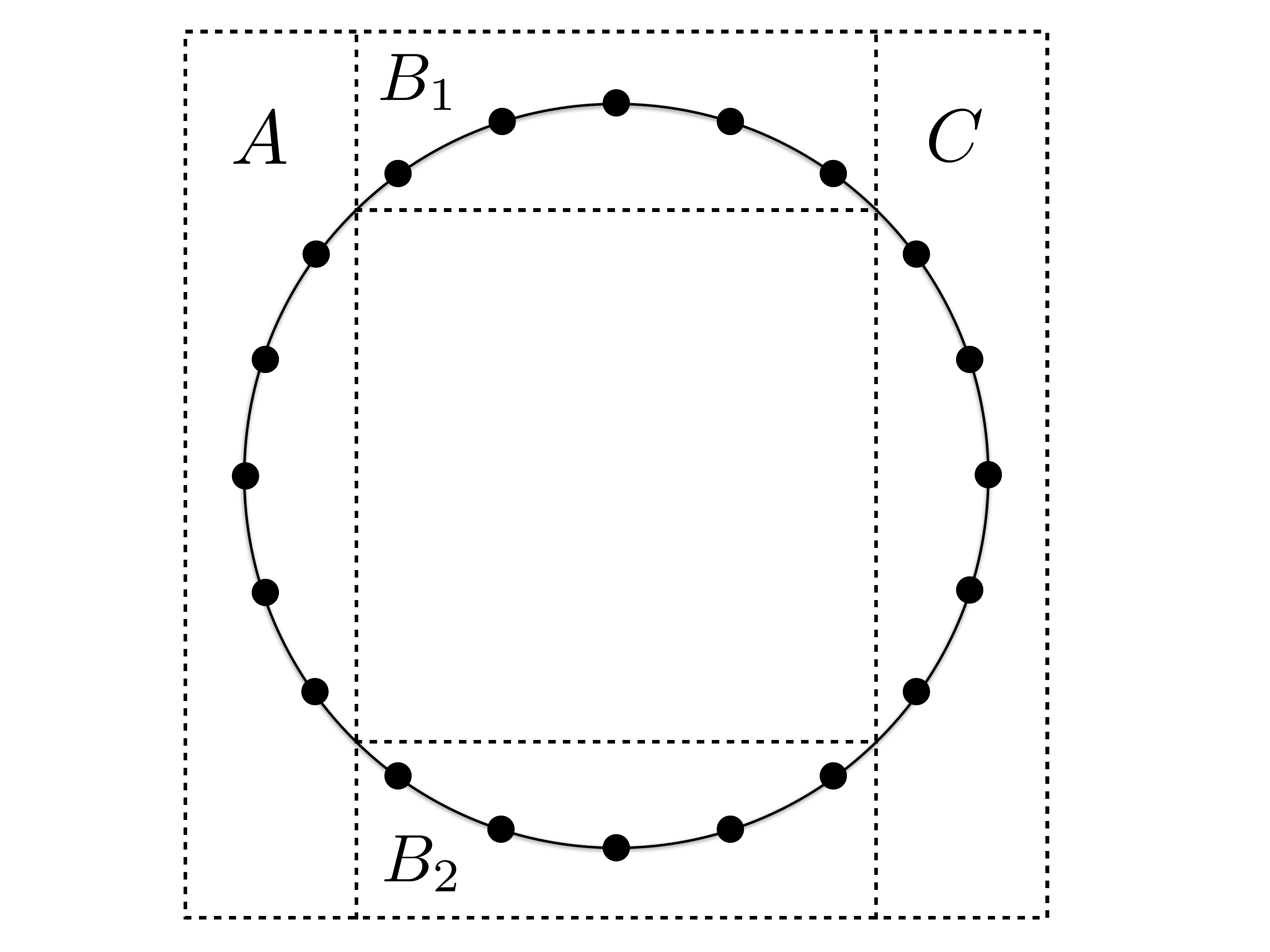}
  \caption{Each dot represents a particle. The partition of a chain to
    three parts $ABC$, where $A,C$ are disconnected and
    $B=B_1\cup B_2$.}
  \label{fig:ring}
\end{figure}

We now illustrate the mutual information approach in one-dimensional
systems. First, consider a one-dimensional system with periodic
boundary conditions. As we need $A$ and $C$ to be large regions far
away from each other, the partition $A,B,C$ can be chosen as in
Fig.~\ref{fig:ring}.

\begin{figure}[ht]
  \centering
  \includegraphics[scale=0.3]{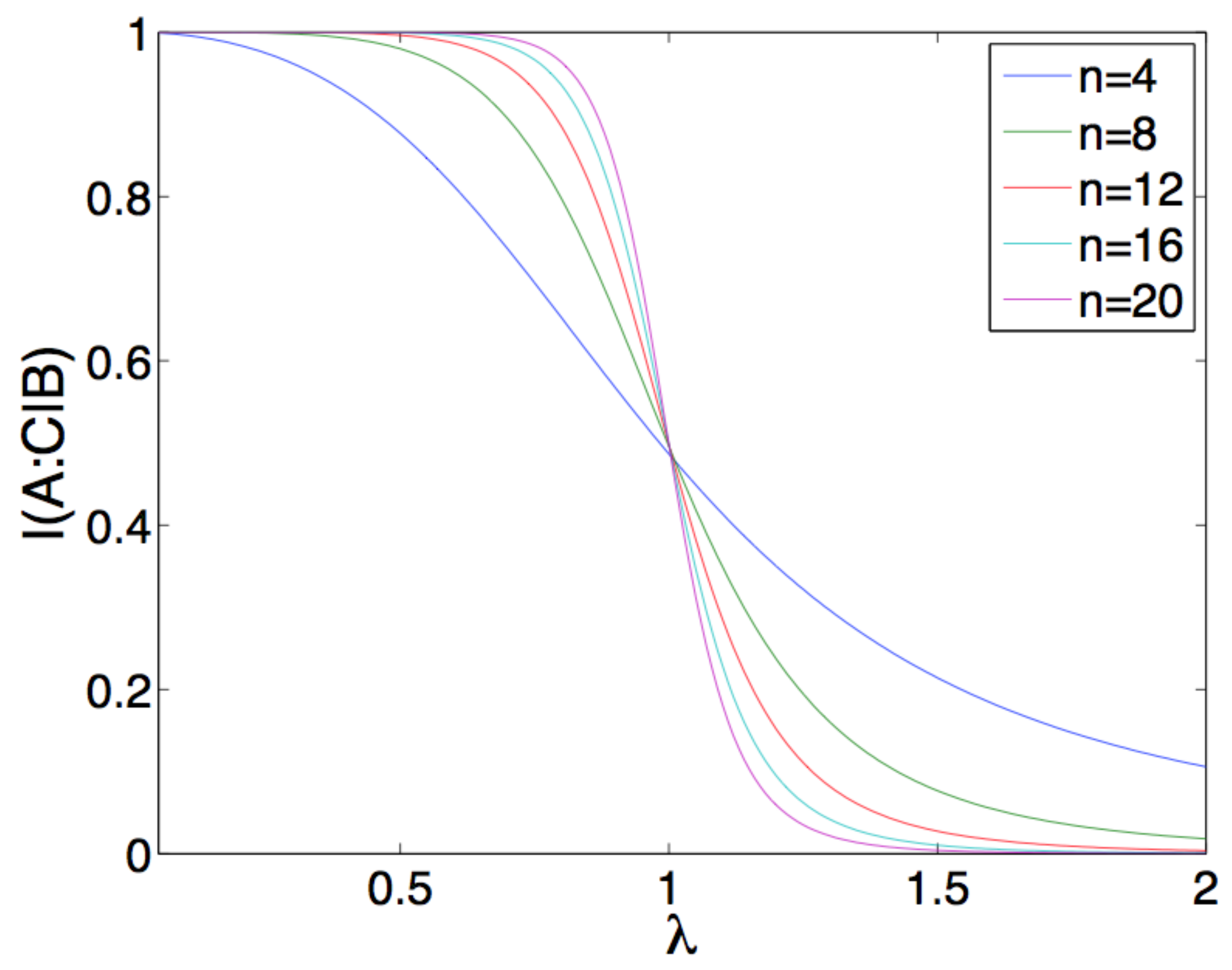}
  \caption{$I(A\,{:}\,C|B)$ of the Ising model with open periodic
    boundary condition and the $A,B,C$ regions as chosen in
    Fig.~\ref{fig:ring}. A similar result is presented in~\cite{LIT},
    from a different viewpoint.}
  \label{fig:IsingT}
\end{figure}

Following the discussions in Sec.~\ref{sec:corr}, one can use the
quantity $I(A\,{:}\,C|B)$ to indirectly detect the existence of the
discontinuity of $\rho^*$ and the corresponding phase transition. We
have computed $I(A\,{:}\,C|B)$ for the ground state of the transverse
quantum Ising chain $H(\lambda)$, with total $4,8,12,16,20$ particles
of the system. The results are shown in Fig.~\ref{fig:IsingT}, in
which $I(A\,{:}\,C|B)$'s clearly indicate a phase transition at
$\lambda=1$ (where the curves intersect). This is consistent with our
discussions for the quantum Ising chain with transverse field in
Sec.~\ref{sec:tqic}.

However, the phase transition of the Hamiltonian with a $Z$ direction
magnetic field, given by
\begin{equation}
  H(\lambda_z)=-J(\sum_{i}Z_iZ_{i+1}+\lambda_z\sum_i Z_i),
\end{equation}
is a \textit{local\/} transition without discontinuity of
$\rho^*(\lambda_z)$. That is, when approached on the boundary of
$\D^{(k)}$, from the direction corresponding to
$\lambda_z\rightarrow 0$, the local RDMs of $\rho^*(\lambda_z)$ has a
sudden change at the point $\lambda_z=0$ (and significantly different
for any two points each corresponding to $\lambda_z<0$ and
$\lambda_z>0$). If we plot the diagram of $I(A\,{:}\,C|B)$ for this
model, we won't see any transition in the system.

We emphasize that the above approach employs calculations of extremely
small systems yet still precisely signals the transition point of the
corresponding system in the thermodynamic limit. For a simple
comparison, the fidelity approach~\cite{gu2010fidelity} for the same
model involves system size of about a thousand and requires the
knowledge of the analytic solutions of the system.

\subsection{The Choice of Regions A,B,C}

\begin{figure}[ht]
  \centering
  \includegraphics[scale=0.35]{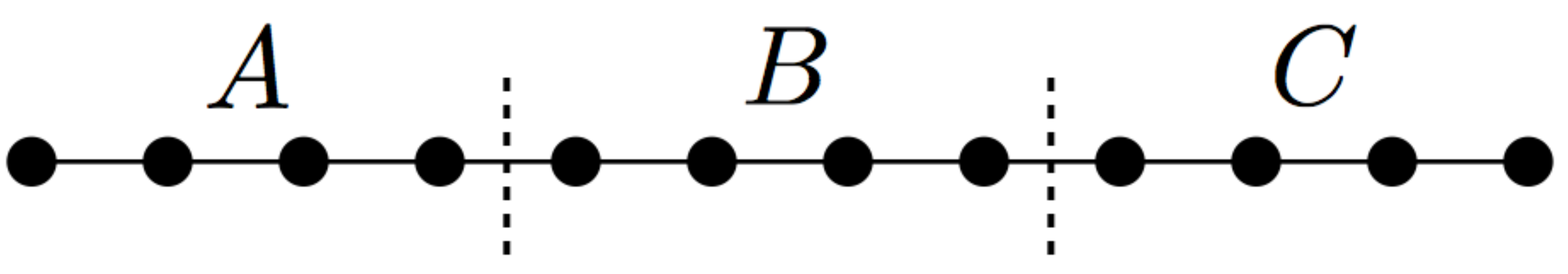}
  \caption{$A,B,C$ cutting on a 1D chain.}
  \label{fig:linepart}
\end{figure}

\begin{figure}[h!]
  \centering
  \includegraphics[scale=0.35]{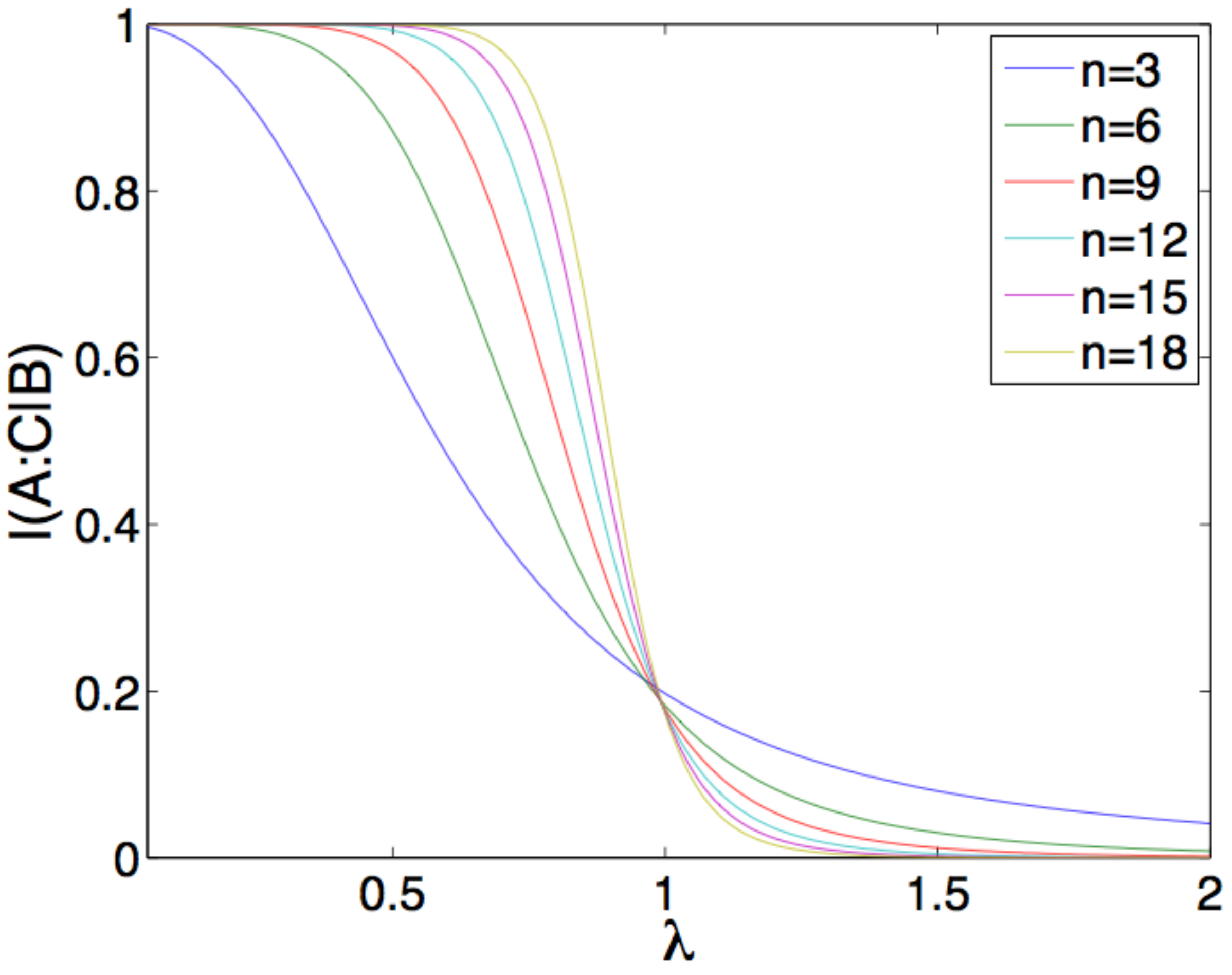}
  \caption{$I(A\,{:}\,C|B)$ of the Ising model with open boundary
    condition and the $A,B,C$ regions as chosen in
    Fig.~\ref{fig:linepart}.}
  \label{fig:IsingOpen}
\end{figure}

It is important to note that the choice of the regions $A,B,C$ should
respect the locality of the system. If we consider one-dimensional
system with open boundary condition, we can choose the $A,B,C$ regions
as in Fig.~\ref{fig:linepart}. For the transverse Ising model with
open boundary condition, this choice will give a similar diagram of
$I(A{:}C|B)$ as in Fig.~\ref{fig:IsingT}, which is given in
Fig.~\ref{fig:IsingOpen}. This clearly shows a discontinuity of
$\rho^*$ and a quantum phase transition at $\lambda=1$.

\begin{figure}[h!]
  \centering
  \includegraphics[scale=0.25]{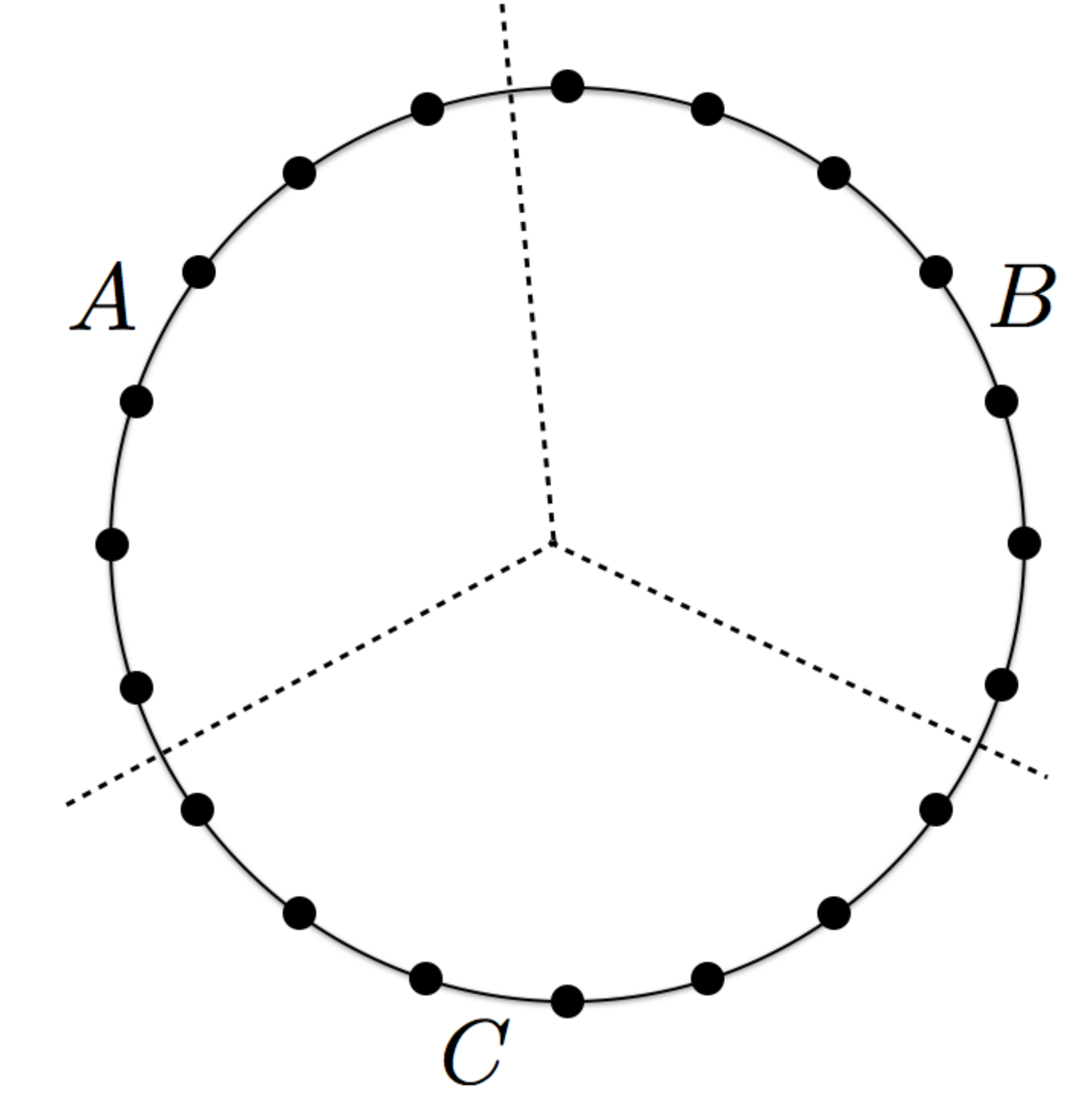}
  \caption{$A,B,C$ cutting on a 1D ring.}
  \label{fig:CuttinRring}
\end{figure}

\begin{figure}[h!]
  \centering
  \includegraphics[scale=0.35]{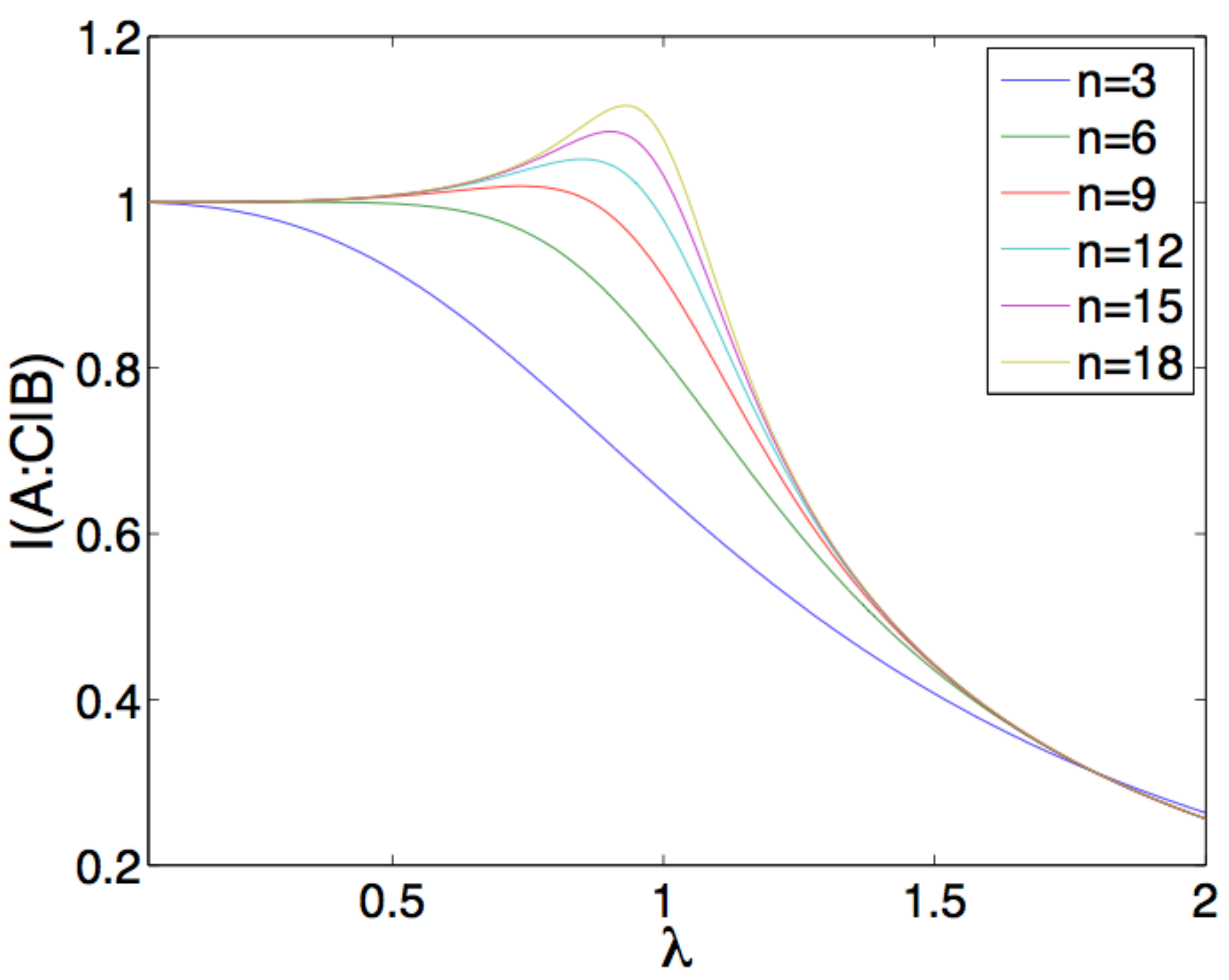}
  \caption{$I(A\,{:}\,C|B)$ of the Ising model with periodical
    boundary condition and the $A,B,C$ regions as chosen in
    Fig.~\ref{fig:CuttinRring}.}
  \label{fig:IsingRing}
\end{figure}

However, if the partition in Fig.~\ref{fig:linepart} is used for the
Ising model with periodical boundary condition, as given in
Fig~\ref{fig:CuttinRring}, the behaviour of $I(A\,{:}\,C|B)$ will be
very different. In fact, in this case $I(A\,{:}\,C|B)$ reflects
nothing but the 1D area law of entanglement, which will diverge at the
critical point $\lambda=1$ in the thermodynamic limit. For a finite
system as illustrated in Fig.~\ref{fig:IsingRing}, $I(A\,{:}\,C|B)$
does no clearly signal the two different quantum phases and the phase
transition.

\subsection{$I(A{:}C|B)$ as a Universal Indicator}

From our previous discussions, we observe that to use $I(A{:}C|B)$ to
detect quantum phase and phase transitions, it is crucial to choose
the areas $A,C$ that are {\it{far from each other}}. Here `far' is
determined by the locality of the system. For instance, on an 1D
chain, the areas $A,C$ in Fig.~\ref{fig:linepart} are far from each
other, but in Fig.~\ref{fig:CuttinRring} are not.

If such an areas $A,C$ are chosen, then for a gapped system, a nonzero
$I(A{:}C|B)$ of a ground state will then indicates non-trial quantum
order. We have already demonstrated it using the transverse Ising
model, where for $0<\lambda<1$, the system exhibits the
`symmetry-breaking' order. In fact, we can also use $I(A{:}C|B)$ to
detect other kind of non-trivial quantum orders.

For instance, $I(A{:}C|B)$ was recently applied to study the quantum
phase transitions related to the so-called `symmetry-protected
topological (SPT) order', which also has a `nonlocal' nature despite
that the corresponding ground states are only short-range entangled
(in the usual sense as discussed in this
paper)~\cite{zeng2014topological}.

\begin{figure}[ht]
  \centering
  \includegraphics[scale=0.35]{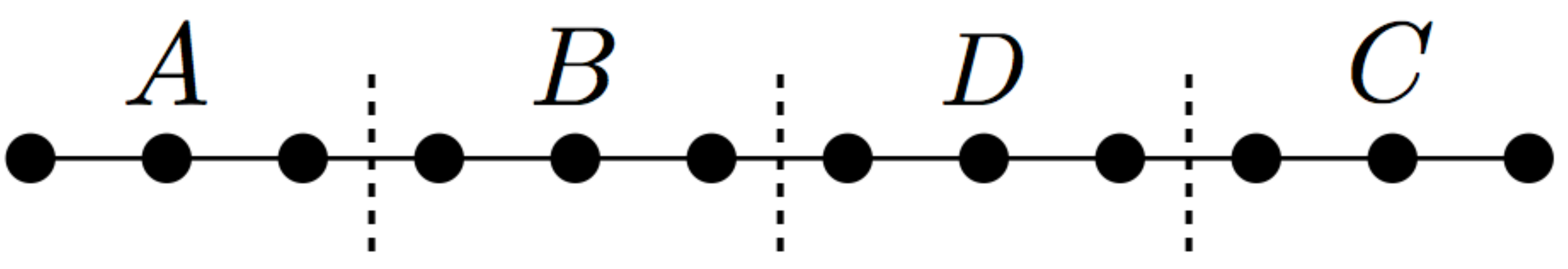}
  \caption{$A,B,C,D$ cutting on a 1D chain}
  \label{fig:line2}
\end{figure}

It was shown that for a 1D gapped system on an open chain, a non-zero
$I(A{:}C|B)$ for the choice of the regions $A,B,C$ as in
Fig.~\ref{fig:linepart} also detects non-trivial SPT order. However,
it does not distinguish SPT order from the symmetry-breaking order. In
stead, one can use a cutting as given in Fig.~\ref{fig:line2}, where
the whole system is divided into four parts $A,B,C,D$, and $I(A{:}C|B)$
the detects the non-trivial correlation in the reduced density matrix
of the state of $ABC$. Under this cutting, $I(A{:}C|B)$ is zero for a
symmetry-breaking ground state, but has non-zero value for an SPT
ground state.


\begin{figure}[ht]
  \centering
  \includegraphics[scale=0.35]{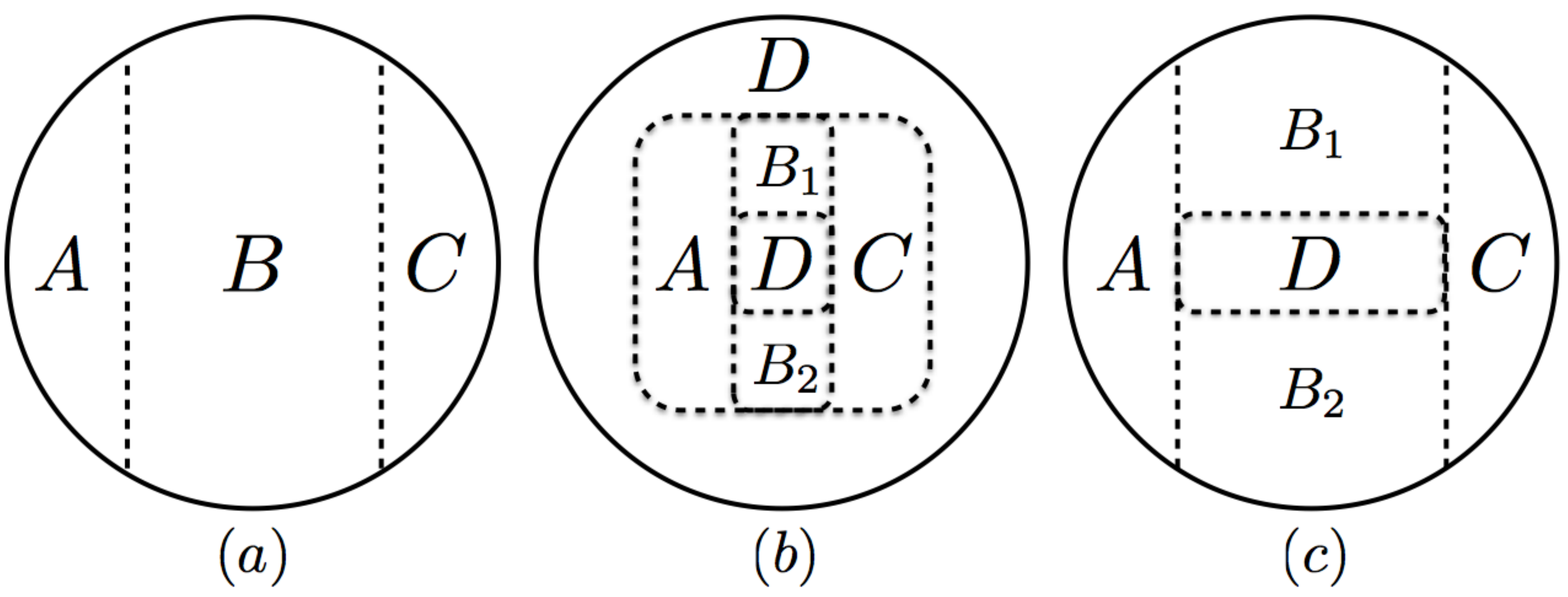}
  \caption{Cuttings on a 2D disk}
  \label{fig:2Dcutting}
\end{figure}

A similar idea also applies to 2D systems. For instance, for a 2D
system on a disk with boundary, we can consider three different kinds
of cuttings~\cite{levin2006detecting,zeng2014topological,LIT}, as
shown in Fig.~\ref{fig:2Dcutting}. For each of these cuttings, a
non-trivial $I(A{:}C|B)$ detects different orders of the system. For
Fig.~\ref{fig:2Dcutting}(a), $I(A{:}C|B)$ detects both symmetry-breaking
order and SPT order and topological phase
transitions~\cite{2014arXiv1402.4245L}. Fig.~\ref{fig:2Dcutting}(b) is
nothing but the choices of $A,B,C$ to define the topological
entanglement entropy by Levin and Wen~\cite{levin2006detecting}, which
detects topological order. And similarly as the 1D case,
Fig.~\ref{fig:2Dcutting}(c) detects SPT order, which distinguishes it
from symmetry-breaking order (in this case $I(A{:}C|B)=0$ for
symmetry-breaking order)~\cite{zeng2014topological}.

In this sense, by choosing proper areas $A,B,C$ with $A,C$ far from
each other, a non-zero $I(A{:}C|B)$ universally indicates a non-trivial
quantum order in the system. Furthermore, by analyzing the choices of
$A,B,C$, it also tells which order the system exhibits
(symmetry-breaking, SPT, topological, or a mixture of them).

We remark that, for a pure state, the cuttings of Fig.~\ref{fig:ring}
and Fig.~\ref{fig:linepart} give that $I(A{:}C|B)=I(A{:}C)$. However, this
is not the case for a mixed state. Therefore, although one may be able
to detect nontrivial quantum order simply using $I(A{:}C)$, in the most
general case, $I(A{:}C|B)$ is a universal indicator of a non-trivial
quantum order but $I(A{:}C)$ is not. For instance, the equal-weight
mixture of the all $\ket{0}$ and all $\ket{1}$ states does not exhibit
non-trivial order (i.e. contains no irreducible many-body
correlation), hence $I(A{:}C|B)=0$ for the cuttings of
Fig.~\ref{fig:ring} and Fig.~\ref{fig:linepart}, but $I(A{:}C)\neq 0$,
which in fact indicates the classical correlation in the system.

\section{Further Properties of the Discontinuity}
\label{sec:further}
In this section, we further explore the structure associated with the
discontinuity of the maximum entropy inference.

\subsection{Path Dependence of Discontinuity}

We continue our discussion of Examples~\ref{eg:numrange1}
to~\ref{eg:numrange3} in dimension $3$, but with more than two
observables. The following example illustrates that one may need to
choose the right path in order to see the discontinuity of $\rho^*$.
It is an example that combines Examples~\ref{eg:numrange1}
and~\ref{eg:numrange2} together.

\begin{example}
  \label{eg:numrange4}
  We consider the tuple $\F$ of $3$ operators, with $F_1,F_2$ the same
  as given in Example~\ref{eg:numrange1} and
  \begin{equation}
    F_3 = \begin{pmatrix} %
      1 & 0 & 1 \\ 0 & 0 & 1 \\ 1 & 1 & -1 %
    \end{pmatrix}.
  \end{equation}
\end{example}

In this example, $\D_\F$ is a compact convex set in $\R^3$. Consider
the point $\balpha=(1,1,0.5)$. If $\balpha$ is approached along the
line $[(1,1,0),(1,1,1)]$, there is no discontinuity of
$\rho^*(\balpha)$, similar as the discussion in
Example~\ref{eg:numrange2}.

However, if $\balpha$ is approached from $\epsilon\rightarrow 0$ in a
Hamiltonian $-F_1+\epsilon F_2$, then there is discontinuity of
$\rho^*(\balpha)$, similar as the discussion in
Example~\ref{eg:numrange1}.

\begin{figure}[ht]
  \centering
  \includegraphics[scale=0.7]{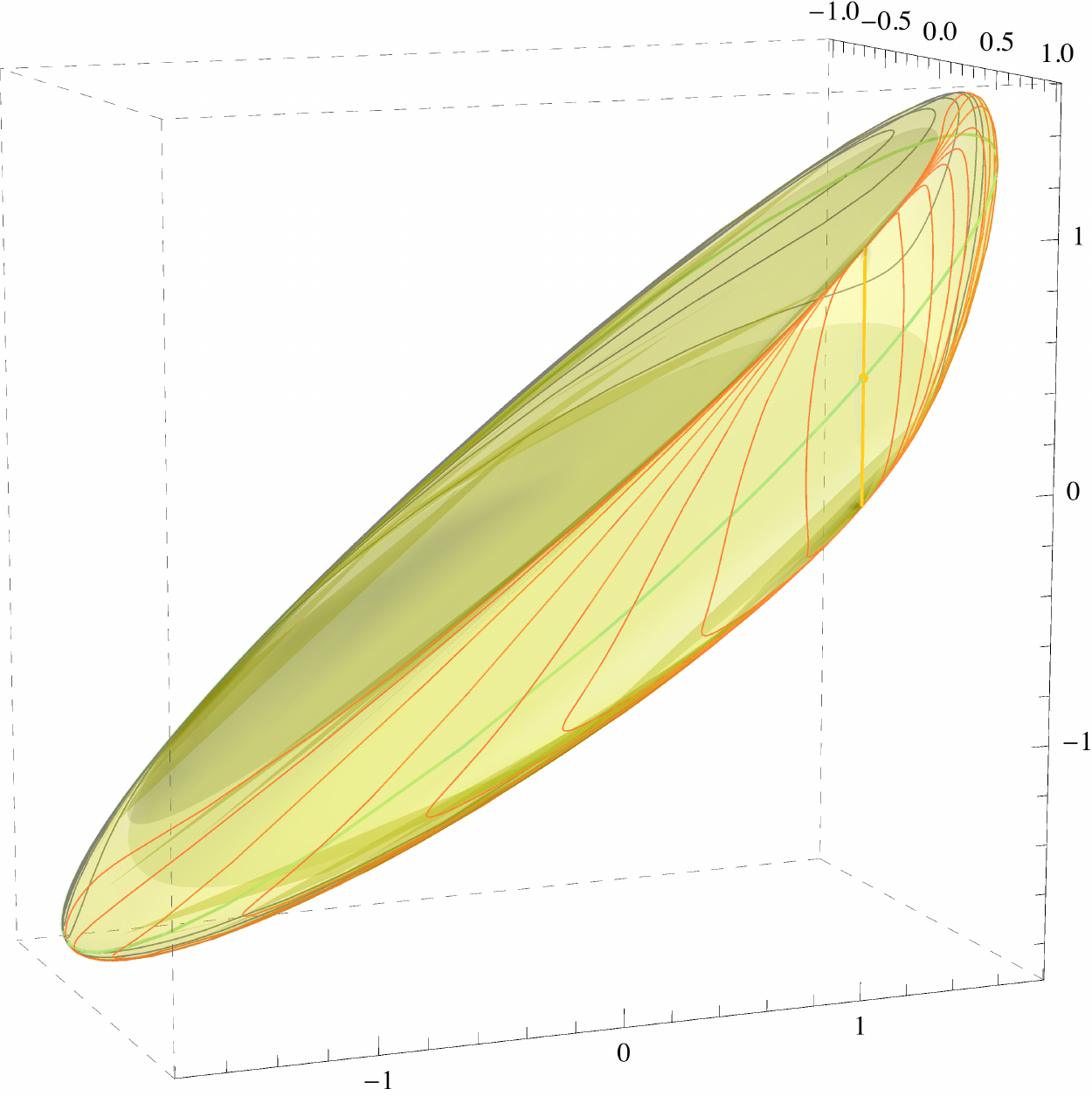}
  \caption{The convex set of $\D_\F$ for $\F=(F_1,F_2,F_3)$ of
    Example~\ref{eg:numrange4} in $\R^3$. For the normalized ground
    state $\rho(\alpha,\phi)$ of
    $\cos\alpha F_1+\sin\alpha\cos\phi F_2+\sin\alpha\sin\phi F_3$ for
    any given $\alpha\in [0,\pi], \phi\in [0,2\pi]$, a point is
    plotted for
    $(\tr(\rho(\alpha,\phi)F_1,\tr(\rho(\alpha,\phi)F_2,\tr(\rho(\alpha,\phi)F_3)$.
    Gray lines correspond to $\alpha\in [0,\pi/2]$, and red lines
    correspond to $\alpha\in [\pi/2,\pi]$. The yellow line corresponds
    to $(1,1,x)$, where the discontinuity happens.}
  \label{fig:Hambondary4}
\end{figure}

The convex set of $\D_\F$ for $\F=(F_1,F_2,F_3)$ in $\R^3$ is shown in
Fig.~\ref{fig:Hambondary4}. This shows that if one approaches the
yellow line (corresponding to $(1,1,x)$) from a line inside the red
area of the surface, then discontinuity of $\rho^*(\balpha)$ happens.
But along the line $[(1,1,0),(1,1,1)]$, there is no discontinuity of
$\rho^*(\balpha)$.

This example shows that, in general for $k$ measurements, whether
there is discontinuity of $\rho^*(\balpha)$ at the point
$\balpha\in\D_\F$ depends on the direction on the boundary of $\D_\F$
along which $\balpha$ is approached. If there is a sequence
$\balpha_s$ approaching $\balpha$ but
\begin{equation}
  \rho^*(\balpha_s)\not\rightarrow
  \rho^*(\balpha),
\end{equation}
then there is discontinuity of $\rho^*(\balpha)$.

The same situation can happen in Example~\ref{eg:GHZ}. If we approach
the $2$-RDM $\brho^{(2)}$ of the GHZ state using the ground states of
$H + \epsilon \sum_{i=1}^3 Z_i$ instead of
$H + \epsilon \sum_{i=1}^3 X_i$ as in Example~\ref{eg:GHZ}, there will
be no discontinuity. And furthermore, for the Hamiltonian
$H+ \epsilon_1 \sum_{i=1}^3 X_i+\epsilon_2 \sum_{i=1}^3 Z_i$, the
convex set of $\D_\F$ for $\F=(F_1,F_2,F_3)$ with
$F_1=Z_1Z_2+Z_2Z_3,F_2= \sum_{i=1}^3 X_i,F_3= \sum_{i=1}^3 Z_i$ has a
similar structure as that in Fig.~\ref{fig:Hambondary4}, as given in
Fig.1c of~\cite{zauner2014symmetry}. Now consider the situation of the
thermodynamic limit, corresponding to the transverse Ising model with
also a magnetic field in the $Z$ direction, i.e. the Hamiltonian
\begin{equation}
  \label{eq:Hxz}
  H(\lambda_x,\lambda_z)=-J(\sum_{i=1}^{n-1} Z_iZ_{i+1} + \lambda_x
  \sum_{i=1}^n X_i+\lambda_z \sum_{i=1}^n Z_i), 
\end{equation}
with $J>0$. The corresponding convex set of $\D_\F$ for
$\F=(F_1,F_2,F_3)$ with
$F_1=\frac{1}{n-1}\sum_{i=1}^{n-1}Z_iZ_{i+1},
F_2=\frac{1}{n}\sum_{i=1}^n X_i, F_3=\frac{1}{n}\sum_{i=1}^n Z_i$
is quite different, as the line of discontinuity (similar as the line
$(1,1,x)$ in Fig.~\ref{fig:Hambondary4}) will expand to become a
`ruled surface' (see Fig.1b of~\cite{zauner2014symmetry}), which is
nothing but the symmetry-breaking phase~\cite{zauner2014symmetry}
(this corresponds to the phase transition at $\lambda=1$).

Another interesting thing of Example~\ref{eg:numrange4} is that the
discontinuities of $\rho^*(\balpha)$ do not only happen at the point
$\balpha=(1,1,0.5)$. In fact they can happen at any point $(1,1,s)$
with $(0<s<1)$. This can be done by engineering the Hamiltonian
\begin{equation}
  H=-F_1+\epsilon F_2+f(\epsilon) F_3,
\end{equation}
with
$\lim\limits_{\epsilon\rightarrow 0}\frac{f(\epsilon)}{\epsilon}=0$
for some function $f(\epsilon)$. We remark that however, this does not
happen in a similar situation of thermodynamic limit. For instance,
the Hamiltonian $H(\lambda_x,\lambda_z)$ discussed above only has one
phase transition (discontinuity) point for $\lambda>0$ at
($\lambda=1$) that corresponds to zero magnetic filed in the $Z$
direction (see Fig.1b of~\cite{zauner2014symmetry}).

\subsection{A Necessary Condition}

Suppose $\rho^*(\balpha_s)\rightarrow\tilde{\rho}$ when
$\balpha_s\rightarrow\balpha$, then we must have
$\tilde{\rho}\in\L(\balpha)$. That is, $\tilde{\rho}$ returns the
measurement results $\balpha$. If discontinuity happens at $\balpha$,
state $\tilde{\rho}$ is different from $\rho^*(\balpha)$. As the
maximal entropy inference $\rho^*$ has the largest range, the range of
$\tilde{\rho}$ is contained in that of $\rho^*$. We can then choose a
linear combination of $\rho^*$ and $\tilde{\rho}$ in $\L(\balpha)$
that has strictly smaller range than $\rho^*$. This then gives us a
necessary condition for discontinuity of $\rho^*(\balpha)$ in finite
dimensions. We emphasize, however, that the same claim may not hold in
infinite systems.

\begin{observation}
  \label{ob:nec}
  A necessary condition for the discontinuity of $\rho^*(\balpha)$ at
  the point $\balpha$ is that there exists a state
  $\tilde{\rho}\in\L(\balpha)$ whose range is strictly contained in
  that of $\rho^*(\balpha)$.
\end{observation}

In particular, for local measurements, we have
\begin{obsprimed}{ob:nec}
  \label{ob:necb}
  A necessary condition for the discontinuity of $\rho^*(\brho^{(k)})$
  at the point $\brho^{(k)}$ is that there exists a state
  $\tilde{\rho}\in\L(\brho^{(k)})$ whose range is strictly contained
  in that of $\rho^*(\brho^{(k)})$.
\end{obsprimed}

To better understand Observation~\ref{ob:necb}, we would like to
examine an example where the condition is not satisfied.

\begin{example}
  \label{eg:sym}
  Consider again a three-qubit system, and the Hamiltonian
  \begin{equation}
    H=H_{12}+H_{23}
  \end{equation}
  as discussed in~\cite{chen2013symmetric}, where $H_{ij}$ acting
  nontrivially on qubits $i,j$ with the matrix form
  \begin{equation}
    \begin{pmatrix}
      \frac{2}{9}&0&0&-\frac{4}{9}\\
      0&\frac{2}{3}&0&0\\
      0&0&\frac{2}{3}&0\\
      -\frac{4}{9}&0&0&\frac{2}{9}
    \end{pmatrix}.
  \end{equation}

  The ground-state space of the Hamiltonian $H$ is two-fold degenerate
  and is spanned by
  \begin{align*}
    \ket{\psi_0} &=\frac{1}{\sqrt{6}}
                   \left(2\ket{000}+\ket{101}+\ket{110}\right),\\
    \ket{\psi_1} &=\frac{1}{\sqrt{6}}
                   \left(2\ket{111}+\ket{010}+\ket{001}\right).
  \end{align*}

  Now take the maximally mixed state
  \begin{equation}
    \rho^{*}=\frac{1}{2}(\ket{\psi_0}\bra{\psi_0}+\ket{\psi_1}\bra{\psi_1}),
  \end{equation}
  and its $2$-RDMs be $\brho^{(2)}$.

  It is straightforward to check that there does not exist any rank $1$
  state in the ground-state space with the form
  $\alpha\ket{\psi_0}+\beta\ket{\psi_1}$ that has the same $2$-RDMs as
  $\brho^{(2)}$. Therefore, for $\rho^{*}(\brho^{(2)})$, the condition
  in Observation~\ref{ob:necb} is not satisfied, hence no discontinuity
  at the point $\brho^{(2)}$.
\end{example}

In the previous subsection, we see that discontinuity of
$\rho^*(\balpha)$ at the point $\balpha\in\D_\F$ depends on the
direction approaching $\balpha$. The next example tells us that one
cannot conclude the existence of discontinuity by looking at the low
dimensional projections of $\D_\F$.

\begin{example}
  \label{eg:numrange5}
  Consider the measurement of $4$ operators, with $F_1,F_2,F_3$ the
  same as given in Example~\ref{eg:numrange4} and
  \begin{equation}
    F_4 = \begin{pmatrix} %
      1 & 1 & 0 \\ 1 & 1 & 0 \\ 0 & 0 & -1 %
    \end{pmatrix}.
  \end{equation}
  And let $\F=(F_1,F_2,F_3,F_4)$.
\end{example}

Note that the projection of $\D_\F$ to the plane spanned by
$(F_1,F_2)$ is nothing but Fig.~\ref{fig:numrange1}, whose maximum
entropy inference has discontinuity at the point $(1,1)$. However, for
the measurements $\F$, one cannot conclude the existence of points of
discontinuity by solely examining the discontinuity at its projections
(e.g. the discontinuity for measuring $(F_1,F_2)$ only). The existence
of $(F_3,F_4)$ does matter.

To see why, for the point $\balpha=(1,1,0.5,1)$, the maximum entropy
inference is
$\rho^*(\balpha) = \frac{1}{2} (\ket{0}\bra{0} + \ket{1}\bra{1})$.
However, there is no rank one state of the form
$\alpha\ket{0}+\beta\ket{1}$ with $|\alpha|^2+|\beta|^2=1$ that can
return the measurement result $\balpha$. Then according to
Observation~\ref{ob:nec}, there is in fact no discontinuity at
$\balpha$.

\subsection{A Sufficient Condition}

Notice that the condition in Observation~\ref{ob:nec} is not
sufficient. Example~\ref{eg:numrange2} provides a counterexample. By
studying the examples that do have discontinuity, we find a sufficient
condition for the discontinuity of $\rho^*$.

\begin{observation}
  \label{ob:suf}
  For a set of observables $\F=(F_1,\ldots,F_r)$, if:
  \begin{itemize}
  \item the ground state space $V_0$ of some Hamiltonian
    $H_0=\sum_{i=1}^r c_iF_i$ is degenerate with the maximally mixed
    state supported on $V_0$ be $\rho^{*}$, which corresponds to
    measurement results $\alpha_i=\tr(\rho^{*}F_i)$;
  \item there exists a basis $\ket{\psi_a}$ of $V_0$ such that
    \begin{equation}
      \label{eq:errdet}
      \bra{{\psi_a}}F_{i}\ket{\psi_b}=\delta_{ab}
    \end{equation}
    for any $a\neq b$ and $F_i\in \F$;
  \item there exists a sequence of
    \begin{equation}
      \boldsymbol{\epsilon}=(\epsilon_1,\ldots,\epsilon_r)\rightarrow
      (0,\ldots,0),
    \end{equation}
    such that the Hamiltonian $H=H_0+\sum_{i=1}^r \epsilon_iF_i$ has
    unique ground states $\ket{\psi(\boldsymbol{\epsilon})}$ at any
    nonzero $\boldsymbol{\epsilon}$, and
    \begin{equation}
      \label{eq:limits}
      \lim\limits_{\boldsymbol{\epsilon}\rightarrow(0,\ldots,0)}\ket{
        \psi(\boldsymbol{\epsilon})}=\ket{\psi},
    \end{equation}
    where $ \ket{\psi}=\frac{1}{\sqrt{m}}\sum_{a=1}^{m}\ket{\psi_a} $ and
    $m$ is the ground state degeneracy of $H_0$ ($m>1$);
  \end{itemize}
  then $\rho^*(\balpha)$ is discontinuous at the point $\balpha$.
\end{observation}

This condition guarantees that the state $\ket{\psi}$ and the
maximally mixed state $\rho^*$ have the same local density matrices.
The discontinuity of maximum entropy inference therefore follows when
considering the sequence of reduced density matrices of
$\ket{\psi(\boldsymbol{\epsilon})}$.

Notice that Eq.~\eqref{eq:errdet} is the quantum error-detecting
condition for the error set $\F$ but without the coherence condition
of $\bra{{\psi_a}}F_{j}\ket{\psi_a}=c_j$ for
$a=b$~\cite{knill1997theory}, where $c_j$ is a constant that is
independent of $a$. We will refer to this condition as the partial
error-detecting condition.

For example, for the observables $\F=(F_1,F_2,F_3)$ discussed in
Example 6, consider the ground-state space of $H_0=-F_1$, which is
degenerate and is spanned by $\{\ket{0},\ket{1}\}$. It is
straightforward to check that $\bra{0}F_i\ket{1}=0$ for all $i=1,2,3$.
Furthermore, the Hamiltonian $H=-F_1+\epsilon_1 F_2+\epsilon_2 F_3$
has a unique ground state $\ket{\psi(\boldsymbol{\epsilon})}$ at any
nonzero $\boldsymbol{\epsilon}=(\epsilon_1,\epsilon_2)\neq 0$. And for
the sequence that $\epsilon_2=0$ and $\epsilon_1\rightarrow 0$,
$ \lim\limits_{\epsilon_1\rightarrow 0}\ket{
  \psi(\epsilon_1,0)}=\frac{1}{\sqrt{2}}(\ket{0}+\ket{1})$.

Similarly for local measurements, we have
\begin{obsprimed}{ob:suf}
  \label{ob:sufb}
  For a set of $k$-local observables $\F=(F_1,\ldots,F_r)$, if:
  \begin{itemize}
  \item the ground state space $V_0$ of some Hamiltonian
    $H_0=\sum_{i=1}^r c_iF_i$ is degenerate with the maximally mixed
    state supported on $V_0$ be $\rho^{*}$, which corresponds to
    $k$-RDMs $\brho^{(k)}$;
  \item there exists a basis $\ket{\psi_a}$ of $V_0$ such that
    \begin{equation}
      \label{eq:errdetl}
      \bra{{\psi_a}}F_{i}\ket{\psi_b}=\delta_{ab}
    \end{equation}
    for any $a\neq b$ and $F_i\in \F$;
  \item there exists a sequence of
    \begin{equation}
      \boldsymbol{\epsilon}=(\epsilon_1,\ldots,\epsilon_r)\rightarrow
      (0,\ldots,0),
    \end{equation}
    such that the Hamiltonian $H=H_0+\sum_{i=1}^r \epsilon_iF_i$ has
    unique ground states $\ket{\psi(\boldsymbol{\epsilon})}$ at any
    nonzero $\boldsymbol{\epsilon}$, and
    \begin{equation}
      \lim\limits_{\boldsymbol{\epsilon}\rightarrow(0,\ldots,0)}
      \ket{\psi(\boldsymbol{\epsilon})}=\ket{\psi},
    \end{equation}
    where $ \ket{\psi}=\frac{1}{\sqrt{m}}\sum_{a=1}^{m}\ket{\psi_a} $ and
    $m$ is the ground state degeneracy of $H_0$ ($m>1$);
  \end{itemize}
  then $\rho^*(\brho^{(k)})$ is discontinuous at the point
  $\brho^{(k)}$.
\end{obsprimed}

For example, for the observables $\F=(F_1,F_2,F_3)$ with
$F_1=Z_1Z_2+Z_2Z_3,F_2= \sum_{i=1}^3 X_i,F_3= \sum_{i=1}^3 Z_i$
discussed in Example 4, consider the ground-state space of $H_0=-F_1$,
which is degenerate and is spanned by $\{\ket{000},\ket{111}\}$. It is
straightforward to check that $\bra{000}F_i\ket{111}=0$ for all
$i=1,2,3$. Furthermore, the Hamiltonian
$H=-F_1+\epsilon_1 F_2+\epsilon_2 F_3$ has a unique ground state
$\ket{\psi(\boldsymbol{\epsilon})}$ at any nonzero
$\boldsymbol{\epsilon}=(\epsilon_1,\epsilon_2)\neq 0$. And for the
sequence that $\epsilon_2=0$ and $\epsilon_1\rightarrow 0$,
$ \lim\limits_{\epsilon_1\rightarrow 0}\ket{
  \psi(\epsilon_1,0)}=\frac{1}{\sqrt{2}}(\ket{000}+\ket{111})$.

These demonstrate an intimate connection between the discontinuity of
$\rho^*(\brho^{(k)})$ and the (partial) quantum error-detecting
condition.

\section{Summary and Discussion}
\label{sec:summary}

We now summarize the main results this paper in Table~\ref{Table1}. We
start from introducing two natural types of quantum phase transitions:
a local type that can be detected by a non-smooth change of local
observable measurements, and a non-local type which cannot. We then
further show that the discontinuity the maximum entropy inference
$\rho^*(\brho^{(k)})$ detects the non-local type of transitions. We
have done this by examining the convex set $\D^{(k)}$ of the local
reduced density matrices $\brho^{(k)}$, where the discontinuity of
$\rho^*(\brho^{(k)})$ only happens on the boundary of the convex set,
hence is directly related to the ground states of local Hamiltonians
(hence zero temperature physics). And essentially, the discontinuity
only happens at the transition points.

We further show that the discontinuity of $\rho^*(\brho^{(k)})$ is in
fact related to the existence of irreducible many-body correlations.
This allows us to propose a practical method for detecting the
non-local type of transitions by the quantum conditional mutual
information of two disconnected parts, which is an analogy of the
Levin-Wen topological entanglement entropy~\cite{levin2006detecting}.
We have demonstrated how the conditional mutual information detects
the phase transition in the transverse Ising model and the toric code
model, which are both continuous quantum phase transitions.

\begin{table}[h!]
  \centering
  \begin{tabular}{p{5cm}||c|c}
    \hline
    Types of Quantum Phase Transitions & Local & Non-Local \\
    \hline
    Discontinuity of $\rho^*(\brho^{(k)})$ & No & Yes \\
    \hline
    Irreducible Many-Body Correlations & No & Yes \\
    \hline
    Conditional Mutual Information & Zero & Nonzero \\
    \hline
  \end{tabular}
  \caption{Summary of the relationship between the main concepts
    discussed in this paper.}
  \label{Table1}
\end{table}

Based on the connection between irreducible many-body correlation and
the quantum conditional mutual information $I(A{:}C|B)$, we have
proposed that $I(A{:}C|B)$ as a universal indicator of non-trivial
quantum order of gapped systems. The crucial part is to chose that the
areas $A,C$ that are far from each other, based on the locality of the
system. By choosing proper regions to compute $I(A\,{:}\,C|B)$, one
can indeed further tell the type of the phase transition
(symmetry-breaking, topological, SPT, or a mixture of them). We
summarize these different indicators in Table~\ref{Table2}.

\begin{table}[h!]
  \centering
  \begin{tabular}{p{5cm}||c|c|c}
    \hline
    & Fig.~\ref{fig:ring} or \ref{fig:linepart} or
      \ref{fig:2Dcutting}(a) & Fig.~\ref{fig:2Dcutting}(b) 
    & Fig.~\ref{fig:line2} or \ref{fig:2Dcutting}(c)\\
    \hline
    symmetry-breaking order & Yes & No & No\\
    \hline
    topological order & No & Yes & No \\
    \hline
    SPT order & Yes & No & Yes \\
    \hline
  \end{tabular}
  \caption{Summary of the choices of the areas of $A,B,C$ (in
    different figures) and the non-trivial indicator $I(A{:}C|B)$ 
    for different quantum order. Here `Yes' means a non-zero value of
    $I(A{:}C|B)$.} 
  \label{Table2}
\end{table}

We remark that a non-zero $I(A{:}C|B)$ even contains information for a
gapless system. By choosing different ratios of the lengths (areas) of
$A,B,C$, the value $I(A{:}C|B)$ of a gapless system could vary, and the
dependance of $I(A{:}C|B)$ with that ratios is closely related to
universal quantities of the system, such as the central
charge~\cite{LIT}.

We hope that our discussions brings new links between quantum
information theory and condensed matter physics.

\section*{Acknowledgments}

We thank Stephan Weis and Xiao-Gang Wen for helpful discussions. JC,
ZJ and NY acknowledge the hospitality of UTS--AMSS Joint Research
Laboratory for Quantum Computation and Quantum Information Processing
where parts of the work were done. ZJ acknowledges support from NSERC
and ARO. YS, NY and BZ are supported by NSERC. DLZ is supported by NSF
of China under Grant No. 11175247 and 11475254, and NKBRSF of China
under Grants Nos. 2012CB922104 and 2014CB921202.

\bibliographystyle{plain}

\bibliography{MaxEntQFT}


\end{document}